\documentclass[12pt]{article}
\input psfig.tex
\usepackage{subeqn}

\newsavebox{\astrutbox}
\sbox{\astrutbox}{\rule[-5pt]{0pt}{20pt}}
\newcommand{\be}{\begin{equation}}
\newcommand{\ee}{\end{equation}}
\def\pa{\parallel}
\def\kpa{k_{\parallel}}
\def\kpn{k_{\perp}}
\def\bB{{\bf B}}
\def\kk{{\bf k}}
\def\pp{{\bf p}}
\def\qq{{\bf q}}
\def\kkpr{{\bf k}^{\prime}}
\def\ep{{\bf \hat{e}}_{\parallel}}
\def\ek{{\bf \hat{e}}_k}
\def\eep{{\bf \hat{e}}_p}

\def\nn{\bf \hat{n}}
\def\ete{{\bf \hat{e}}_{\theta}}
\def\efi{{\bf \hat{e}}_{\Phi}}
\def\kkp{{\bf k}_{\perp}}
\def\ols{\omega_{\Lambda}^s}
\def\olsp{\omega_{\Lambda_p}^{s_p}}
\def\olsq{\omega_{\Lambda_q}^{s_q}}
\def\hh{{\bf h^{\Lambda}_k}}
\def\Lp{\Lambda^{\prime}}

\def\aak{a_{\Lambda}^s}
\def\aap{a_{\Lambda_p}^{s_p}}
\def\aaq{a_{\Lambda_q}^{s_q}}
\def\aakp{a_{\Lambda^{\prime}}^{s^{\prime}}}
\def\aakdp{a_{\Lambda^{\prime \prime}}^{s^{\prime \prime}}}
\def\qls{q_{\Lambda}^s}
\def\qlsprime{q_{\Lambda^{\prime}}^{s^{\prime}}}
\def\qlsdprime{q_{\Lambda^{\prime \prime}}^{s^{\prime \prime}}}
\def\qlsp{q_{\Lambda_p}^{s_p}}
\def\qlsq{q_{\Lambda_q}^{s_q}}
\def\alf{\xi_{\Lambda}^s}
\def\alfp{\xi_{\Lambda_p}^{s_p}}
\def\alfq{\xi_{\Lambda_q}^{s_q}}
\def\alfm{\xi_{\Lambda}^{-s}}
\def\alfmp{\xi_{\Lambda_p}^{-s_p}}
\def\alfmq{\xi_{\Lambda_q}^{-s_q}}
\def\alfprime{\xi_{\Lambda^{\prime}}^{s^{\prime}}}
\def\alfmprime{\xi_{\Lambda^{\prime}}^{-s^{\prime}}}
\def\alfdprime{\xi_{\Lambda^{\prime \prime}}^{s^{\prime \prime}}}
\def\alfmdprime{\xi_{\Lambda^{\prime \prime}}^{-s^{\prime \prime}}}
\def\eg{{\it e.g.}\ }

\def\ie{{\it i.e.}\ }

\mathchardef\varLambda="0103

\begin{document}
\begin{center}
{\bf Wave turbulence in incompressible Hall magnetohydrodynamics} \\
S. Galtier
\vskip0.4truein
Institut d'Astrophysique Spatiale (IAS), B\^atiment 121, F-91405 Orsay 
(France); Universit\'e Paris-Sud 11 and CNRS (UMR 8617)\\

\vskip0.5truein
{\it J. Plasma Phys., in press}
\end{center}
\vskip0.5truein

\begin{abstract}
We investigate the steepening of the magnetic fluctuation power law spectra 
observed in the inner solar wind for frequencies higher than $0.5$~Hz. This high 
frequency part of the spectrum may be attributed to dispersive nonlinear 
processes. In that context, the long-time behavior of 
weakly interacting waves is examined in the framework of three--dimensional 
incompressible Hall magnetohydrodynamic (MHD) turbulence. The Hall term added to the 
standard MHD equations makes the Alfv\'en waves dispersive and circularly polarized. 
We introduce the generalized Els\"asser variables and, using a complex helicity 
decomposition, we derive for three-wave interaction processes the general wave 
kinetic equations; they describe the nonlinear dynamics of Alfv\'en, whistler 
and ion cyclotron wave turbulence in the presence of a strong uniform 
magnetic field $B_0 \ep$. Hall MHD turbulence is characterized by anisotropies of 
different strength: (i) for wavenumbers $k d_i \gg 1$ ($d_i$ is the ion inertial 
length) nonlinear transfers are {\it essentially} in the direction perpendicular 
($\perp$) to ${\bf B_0}$; (ii) for $k d_i \ll 1$ nonlinear transfers are {\it 
exclusively} in the perpendicular direction; (iii) for $k d_i \sim 1$, a moderate 
anisotropy is predicted. We show that electron and standard MHD turbulence can be 
seen as two frequency limits of the present theory but the standard MHD limit is 
singular; additionally, we analyze in detail the ion MHD turbulence 
limit. Exact power law solutions of the master wave kinetic equations are given 
in the small and large scale limits for which we have, respectively, the total 
energy spectra $E(\kpn,\kpa) \sim \kpn^{-5/2} |\kpa|^{-1/2}$ and 
$E(\kpn,\kpa) \sim \kpn^{-2}$. An anisotropic phenomenology is developed to 
describe continuously the different scaling laws of the energy spectrum; 
one predicts $E(\kpn,\kpa) \sim \kpn^{-2} |\kpa|^{-1/2} (1+\kpn^2d_i^2)^{-1/4}$. 
Nonlocal interactions between Alfv\'en, whistler and ion cyclotron waves are 
investigated; a non trivial dynamics exists only when a discrepancy from the 
equipartition between the large scale kinetic and magnetic energies happens. 
\end{abstract}

\section{Introduction}
\label{intro}

Spacecraft observations of the inner solar wind, \ie for heliocentric distances 
less than $1$AU, show magnetic and velocity fluctuations over a broad range of 
frequencies, from $10^{-5}$Hz up to several hundred Hz (see \eg Coleman, 1968; 
Belcher and Davis, 1971; Matthaeus and Goldstein, 1982; Roberts et al., 1987; 
Grappin et al., 1990; Burlaga, 1991; Leamon et al., 1998). These fluctuations 
possess many properties expected of fully developed weakly compressible 
magnetohydrodynamic (MHD) turbulence (Goldstein and Roberts, 1999). For that reason 
the interplanetary medium is often seen as a vast laboratory for studying many 
fundamental questions about turbulent plasmas. 

Compressed interaction regions produced by velocity differences are clearly observed 
in the outer solar wind. In regions where fast streams overtake slow streams, the 
spectrum of the large scale magnetic field fluctuations follows a $f^{-2}$ frequency 
power law (Burlaga et al., 1987) which was shown to be a spectral signature of jumps 
(Roberts et al., 1987). Compressive effects are however weaker in the inner solar wind. 
For example, the normalized density fluctuations near the current sheet at $0.3$AU are 
often smaller than $5$\% (Bavassano et al., 1997). This tendency is confirmed by indirect
measurements through radio wave interplanetary scintillation observations at heliocentric 
distances of $16-26$ solar radius (Spangler, 2002). Waves and turbulence -- the subject 
of this paper -- are better observed in the pure/polar wind where generally the 
density fluctuations are weaker than in the current sheet. Therefore, the inner solar
wind may be seen mainly as a weakly compressible medium. 

The turbulent state of the solar wind was suggested by \cite{Coleman68} who reported 
a power law behavior for energy spectra with spectral indices lying between $-1$ and 
$-2$. The original signals being measured in time, these observational spectra are 
measured in frequency. Since the solar wind is super-sonic and super-Alfv\'enic, the 
Taylor ``frozen-in flow'' hypothesis is usually used to connect directly a frequency 
to a wavenumber which allows eventually a comparison with theoretical predictions. 
Note, however, that for anisotropic turbulence a relevant comparison with theoretical 
predictions, like those made in this paper, are only possible if a three dimensional 
energy spectrum is accessible by {\it in situ} measurements, a situation that is not
currently achieved (see, however, Matthaeus et al., 2005; see also the last Section). 
More precise measurements (see \eg Matthaeus and Goldstein, 1982) 
revealed that the spectral index at low frequency is often about $-1.7$ 
which is closer to the Kolmogorov prediction (Kolmogorov, 1941) for neutral 
fluids ($-5/3$) rather than the Iroshnikov-Kraichnan prediction (Iroshnikov, 
1963; Kraichnan, 1965) for magnetized fluids ($-3/2$). Both predictions are 
built, in particular, on the hypothesis of isotropic turbulence. However, the 
presence of Alfv\'en waves in the fast solar wind attests that the magnetic field 
has a preferential direction which is likely at the origin of anisotropic turbulence 
provided that the amount of counter-propagating Alfv\'en waves is enough. Indeed, 
{\it in situ} measurements of cross helicity show clearly that outward propagative 
Alfv\'en waves are the main component of the fast solar wind at short radial 
distances (Belcher and Davis, 1971) but they become less dominant beyond $1$AU. 
Since pure Alfv\'en waves are exact solutions of the ideal incompressible MHD 
equations (see \eg Pouquet, 1993), nonlinear interactions should be suppressed 
if only one type of waves is present. 
The variance analysis of the magnetic field components and magnitude shows clearly 
that the magnetic field vector of the fast solar wind has a varying direction but 
with only a weak variation in magnitude (see \eg Forsyth et al., 1996ab). Typical 
values give a ratio of the normalized variance of the field magnitude smaller than 
$10$\% whereas for the components it can be as large as $50$\%. In these respects,
the interplanetary magnetic field may be seen as a vector liying approximately 
around the Parker spiral direction with only weak magnitude variations (Barnes, 1981). 
Solar wind anisotropy with more power perpendicular to the mean 
magnetic field than that parallel is pointed out by data analyses (Belcher and Davis, 
1971; Klein et al., 1993) with a ratio of power up to $30$. From single-point spacecraft 
measurements it is however not possible to specify the exact three-dimensional form 
of the spectral tensor of the magnetic or velocity fluctuations. In absence of such 
data, \cite{Bieb} with a quasi two-dimensional model, in which wave vectors are 
nearly perpendicular to the large-scale magnetic field, argued that about $85\%$ of 
solar wind turbulence possesses a dominant 2D component. Additionally, solar wind 
anisotropies is detected through radio wave scintillations which reveal that density 
spectra close to the Sun are highly anisotropic with irregularities stretched out 
mainly along the radial direction (Armstrong et al., 1990). 

Most of the papers dealing with interplanetary turbulence tend to focus on a 
frequency inertial range where the MHD approximation is well satisfied. It is the 
domain where the power spectral index is often found to be around the Kolmogorov
index. During the 
last decades several properties of the solar wind turbulence have been understood in 
this framework. Less understood is what happens outside of this range of frequencies. 
At lower frequencies ($<10^{-5}$Hz) flatter power laws are found in particular for 
the fast solar wind with indices close to $-1$ and even less (in absolute value) for 
the smallest frequencies. These large scales are often interpreted as the energy 
containing scales. But the precise role of the low solar corona in the generation 
of such scales and the origin of the evolution of the spectral index when the 
distance from the Sun increases are still not well understood (see \eg Velli et 
al., 1989; Horbury, 1999). 
For frequencies higher than $0.5$Hz a steepening of the magnetic fluctuation 
power law spectra is observed over more than two decades (Coroniti et al., 1982; 
Denskat et al., 1983, Leamon et al., 1998) with a spectral index on average around 
$-3$. Note that the latest analysis made with the Cluster spacecraft data reveals 
a less steep spectral index about $-2.12$ (Bale et al., 2005). This new range, 
exhibiting a power law, is characterized by a bias of the polarization suggesting 
that these fluctuations are likely to be right-hand polarized, outward propagating 
waves (Goldstein et al., 1994). Various indirect lines of evidence indicate that 
these waves propagate at large angles to the background magnetic field and that the 
power in fluctuations parallel to the background magnetic field is much less than 
the perpendicular one (Coroniti et al., 1982; Leamon et al., 1998). For these 
reasons, it is thought (\eg Stawicki et al., 2001) that Alfv\'en -- left circularly 
polarized -- 
fluctuations are suppressed by proton cyclotron damping and that the high frequency 
power law spectra are likely to consist of whistler waves. This scenario proposed 
is supported by direct numerical simulations of compressible $2{1 \over 2}$D Hall 
MHD turbulence (Ghosh et al., 1996) where a steepening of the spectra is found and 
associated with the appearance of right circularly polarized fluctuations. It is 
plausible that what has been conventionally thought of as a dissipation range is 
actually a dispersive or inertial range and that the steeper power law may be due 
to nonlinear wave processes rather than dissipation (see \eg Krishan and Mahajan, 
2004). Under this new interpretation, the resistive dissipation range of frequencies 
may be moved to frequencies higher than the electron cyclotron frequency. 
A recent study (Stawicki et al., 2001) suggests that the treatment of the solar wind 
dispersive range should include magnetosonic/whistler waves since it is often 
observed that the high frequency fluctuations of the magnetic field are much smaller 
than the background magnetic field (see also Forsyth et al., 1996ab). It seems 
therefore that a nonlinear theory built on weak wave turbulence may be a useful 
point of departure for understanding the detailed physics of solar wind turbulence. 

It is well known that the presence of a mean magnetic field plays a fundamental 
role in the behavior of compressible or incompressible MHD turbulence 
(Montgomery and Turner, 1981; Shebalin et al. 1983; Oughton et al., 1994; 
Matthaeus et al., 1996; Kinney and McWilliams, 1998; Cho and Vishniac, 2000;
Oughton and Matthaeus, 2005). 
The main effect is that the mean magnetic field renders the turbulence 
quasi-bidimensional with a nonlinear transfer essentially perpendicular to its 
direction. Note that in such a situation, it is known that compressible MHD behaves 
very similarly to reduced MHD (see \eg Dmitruk et al, 2005). 
This property is generalized to $2{1 \over 2}$D compressible Hall MHD 
for high and low beta $\beta$ plasma simulations (Ghosh and Goldstein, 1997) for 
which strong anisotropies are also found when a strong mean magnetic field is 
present. It was suggested that the action of the Hall term is to provide additional 
suppression of energy cascades along the mean field direction and incompressibility 
may not be able to reproduce the dynamics seen in the simulations. This conclusion 
is slightly different to what we find in the present study where we show that 
incompressibility in Hall MHD allows to reproduce several properties observed like 
anisotropic turbulence. 
Incompressible Hall MHD is often used to understand, for example, the main impact of 
the Hall term on flowing plasmas (Ohsaki, 2005), or in turbulent dynamo (Mininni et al, 
2003). In the present paper, we derive a weak wave turbulence formalism for incompressible 
Hall MHD in the presence of a strong external mean magnetic field, where Alfv\'en, 
ion cyclotron and whistler/electron waves are taken into account. One of the main 
results is the derivation of the wave kinetic equations at the lowest order, \ie for 
three-wave interaction processes. For such a turbulence, it is possible to show, in 
particular, a global tendency towards anisotropy with nonlinear transfers preferentially 
in the direction perpendicular to the external magnetic field. 
Hall MHD wave turbulence theory describes a wide range of frequencies, from the low 
frequency limit of pure Alfv\'enic turbulence to the high frequency limit of whistler 
wave turbulence for which the asymptotic theories have been derived recently 
(Galtier et al., 2000-2002; Galtier and Bhattacharjee, 2003). By recovering both 
theories as two particular limits, we recover all the well-known properties associated. 
Ion cyclotron wave turbulence appears as a third particular limit for which we report 
a detailed analysis. The energy spectrum of Hall MHD is characterized by two inertial 
ranges, which are exact solutions of the wave kinetic equations, separated by a knee. 
The position of the knee corresponds to the scale where the Hall term becomes 
sub/dominant. We develop a single anisotropic phenomenology that recovers the power 
law solutions found and makes the link continuously in wavenumbers between the two 
scaling laws. A nonlocal analysis performed on the wave kinetic equations reveals 
that a non trivial dynamics between Alfv\'en, whistler and ion cyclotron waves 
happens only when a discrepancy from the equipartition between the large scale kinetic 
and magnetic energies exists. We believe that the description given here may help to 
better understand the inner solar wind observations and, in particular, the existence 
of dispersive/non-dispersive inertial ranges. 

The organization of the paper is as follows: in Section 2, we discuss about the 
approximation of incompressible Hall MHD and the existence of transverse, circularly 
polarized waves; we introduce the generalized Els\"asser variables and the complex 
helical decomposition. In Section 3, we develop the wave turbulence formalism and we 
derive the wave kinetic equations for three-wave interaction processes. Section 4 is 
devoted to the general properties of Hall MHD wave turbulence. Section 5 deals with 
the small and large scale limits respectively of the master equations of wave 
turbulence. In particular, we give a detail study of the ion cyclotron wave turbulence. 
In Section 6, we derive an anisotropic heuristic description which, in particular, makes 
the link between the previous predictions. In Section 7, we analyze nonlocal interactions 
between Alfv\'en, whistler and ion cyclotron waves. Section 8 is devoted to the 
possible sources of anisotropy. In Section 9, we discuss about the domain of validity 
of the theory and the difference between wave and strong turbulence. Finally, we 
conclude with a summary and a general discussion in the last Section.

\section{Hall magnetohydrodynamics approximation}
\label{section2}
\subsection{Generalized Ohm's law}

Hall MHD is an extension of the standard MHD where the ion inertia is retained in Ohm's 
law. The generalized Ohm's law, in SI unit, is then given by
\be
{\bf E} + {\bf V} \times \bB - \frac{{\bf J} \times \bB}{n\,e} = 
\mu_0 \eta \, {\bf J} \, ,
\label{equ1}
\ee
where ${\bf E}$ is the electric field, ${\bf V}$ is the plasma flow velocity, $\bB$ is 
the magnetic field, ${\bf J}$ is the current density, $n$ is the electron density, $e$ 
is the magnitude of the electron charge, $\mu_0$ is the permeability of free space and 
$\eta$ is the magnetic diffusivity. The Hall effect, represented by the last term in the 
left hand side of the generalized Ohm's law, becomes relevant when we intend to describe 
the plasma dynamics up to length scales shorter than the ion inertial length $d_i$ 
($d_i = c / \omega_{pi}$, where $c$ is the speed of light and $\omega_{pi}$ is the ion 
plasma frequency) and time scales of the order, or shorter, than the ion cyclotron period 
$\omega_{ci}^{-1}$. It is one of the most important manifestations of the velocity 
difference between electrons and ions when kinetic effects are not taken into account. 
The importance of the Hall effect in astrophysics has been pointed out to understand, 
for example, the presence of instabilities in protostellar disks (Balbus and Terquem, 
2001), the magnetic field evolution in neutron star crusts (Goldreich and Reisenegger, 
1992; Cumming et al, 2004), impulsive magnetic reconnection (see \eg Bhattacharjee, 
2004) or the formation of filaments (see \eg Passot and Sulem, 2003; Dreher et al., 
2005).

\subsection{Incompressible Hall MHD equations}

The inclusion of the Hall effect in the Ohm's law leads, in the incompressible case, to 
the following Hall MHD equations: 
\be
\nabla \cdot {\bf V} = 0 \, ,
\label{hmhd1}
\ee
\be
\frac{\partial {\bf V}}{\partial t} + {\bf V} \cdot \nabla \, {\bf V} = 
- {\bf \nabla} P_* + \bB \cdot \nabla \, \bB + \nu \nabla^2 {\bf V} \, ,
\label{hmhd2}
\ee
\be
\frac{\partial \bB}{\partial t} + {\bf V} \cdot \nabla \, \bB = 
\bB \cdot \nabla \, {\bf V} - d_i \, \nabla \times [ (\nabla \times \bB) \times \bB ]
+ \eta {\nabla}^2\bB \, ,
\label{hmhd3}
\ee
\be
\nabla \cdot \bB = 0 \, ,
\label{hmhd4}
\ee
where $\bB$ has been normalized to a velocity ($\bB \to \sqrt{\mu_0 n m_i} \, \bB$, with 
$m_i$ the ion mass), $P_*$ is the total (magnetic plus kinetic) pressure and $\nu$ is 
the viscosity. The Hall effect appears in the induction equation as an additional term 
proportional to the ion inertial length $d_i$ which means that it is effective when the 
dynamical scale is small enough. In other words, for large scale phenomena this term is 
negligible and we recover the standard MHD equations. In the opposite limit, \eg for very 
fast time scales ($\ll \omega_{ci}^{-1}$), ions do not have time to follow electrons and 
they provide a static homogeneous background on which electrons move. Such a model where 
the dynamics is entirely governed by electrons is called the Electron MHD (EMHD) model 
(Kingsep et al., 1990; Shukla and Stenflo, 1999). It can be recovered from Hall MHD by 
taking the limits of small velocity ${\bf V}$ and large $d_i$. The Electron and Hall MHD 
approximations are particularly relevant 
in the context of collisionless magnetic reconnection where the diffusion region develops 
multiscale structures corresponding to ion and electron characteristic lengths (Huba, 
1995; Biskamp, 1997). For example, it is often considered that whistler/EMHD turbulence 
may act as a detector for magnetic reconnection at the magnetopause (Cai et al., 2001).

\subsection{Three-dimensional inviscid invariants}
\label{invariants}

The three inviscid ($\nu=\eta=0$) quadratic invariants of incompressible Hall MHD are
the total energy 
\be
E = \frac{1}{2} \int ( {\bf V}^2 + \bB^2 ) \, d\,{\cal V} \, ,
\label{I1}
\ee
the magnetic helicity 
\be
H_m = \frac{1}{2} \int {\bf A} \cdot \bB \, d\,{\cal V} \, ,
\label{I2}
\ee
and the generalized hybrid helicity
\be
H_G = \frac{1}{2} \int ({\bf A} + d_i {\bf V}) \cdot (\bB + d_i \nabla \times {\bf V})
\, d\,{\cal V} \, ,
\label{I3}
\ee
with ${\bf A}$ the vector potential ($\bB = \nabla \times {\bf A}$). The third invariant 
generalizes the cross-helicity, $H_c = (1/2) \int {\bf V} \cdot \bB \, d\,{\cal V}$,
which is not conserved anymore when the Hall term is present in the MHD equations. 
The generalized hybrid helicity can be seen as the product of two generalized quantities, 
a vector potential ${\bf \Upsilon}={\bf A} + d_i {\bf V}$ and a vorticity 
${\bf \Omega}= \bB + d_i \nabla \times {\bf V}$. The role played by the generalized 
vorticity is somewhat equivalent to the one played by the magnetic field in standard 
MHD (Woltjer, 1958). Indeed, both quantities obey the same Lagrangian equation, which 
is for ${\bf \Omega}$,
\be
\frac{d {\bf \Omega}}{dt} = {\bf \Omega} \cdot \nabla \, {\bf V} \, . 
\ee
By applying the Helmholtz's law (see \eg Davidson, 2001) to Hall MHD, we see that the 
generalized vorticity lines are frozen into the plasma (see \eg Sahraoui et al., 2003). 
The presence of the Hall term breaks such a property for the magnetic field which 
is however still frozen but only in the electron flow: the introduction of the electron
velocity ${\bf V_e}$, with 
${\bf V} \times \bB - {\bf J} \times \bB / n\,e \simeq {\bf V_e} \times \bB$, leads to 
\be
\frac{d \bB}{dt} = \bB \cdot \nabla \, {\bf V_e} \, , 
\ee
which proves the statement. We will see below that the detailed conservation of 
invariants is the first test that the wave kinetic equations have to satisfy.

\subsection{Incompressible Hall MHD waves}

One of the main effects produced by the presence of the Hall term is that the linearly 
polarized Alfv\'en waves, solutions of the standard MHD equations, become circularly 
polarized and dispersive (see \eg Mahajan and Krishan, 2005; Sahraoui et al., 2005). 
Indeed, if we linearize equations (\ref{hmhd1})--(\ref{hmhd4}) around a strong uniform 
magnetic field $\bB_0$ such that, 
\be
\bB ({\bf x}) = B_0 \, \ep + \epsilon \, {\bf b} ({\bf x}) \, ,
\ee
\be
{\bf V} ({\bf x}) = \epsilon \, {\bf v} ({\bf x}) \, ,
\ee
with $\epsilon$ a small parameter ($0<\epsilon \ll 1$), ${\bf x}$ a 
three-dimensional displacement vector, and $\ep$ a unit vector ($|\ep|=1$), then 
we obtain the following inviscid equations in Fourier space:
\be
\kk \cdot {\bf v}_\kk = 0 \, ,
\label{wave1}
\ee
\be
\partial_t {\bf v}_\kk - i \kpa B_0 {\bf b}_\kk = 
\epsilon \, \{ - {\bf v} \cdot \nabla \, {\bf v} - {\bf \nabla} P_* + 
{\bf b} \cdot \nabla \, {\bf b} \}_\kk \, ,
\label{wave2}
\ee
\be
\partial_t {\bf b}_\kk - i \kpa B_0 {\bf u}_\kk - d_i B_0 \kpa \, \kk \times {\bf b}_\kk = 
\epsilon \, \{ - {\bf v} \cdot \nabla \, {\bf b} + {\bf b} \cdot \nabla \, {\bf v} 
- d_i \, \nabla \times [ (\nabla \times {\bf b}) \times {\bf b} ] \}_\kk \, ,
\label{wave3}
\ee
\be
\kk \cdot {\bf b}_\kk = 0 \, ,
\label{wave4}
\ee
where the wavevector $\kk = k\ek = \kkp + \kpa \ep$ ($k=|\kk|$, $\kpn = |\kkp|$, 
$|\ek|=1$) and $i^2 = -1$. The index $\kk$ denotes the Fourier transform, defined by 
the relation
\be
{\bf v} ({\bf x}) \equiv \int {\bf v} (\kk) \, e^{i \kk \cdot {\bf x}} \, d\kk \, ,
\ee
where ${\bf v} (\kk) = {\bf v}_\kk = {\tilde {\bf v}}_\kk e^{-i \omega t}$ 
(the same notation is used for the magnetic 
field). The linear dispersion relation ($\epsilon=0$) reads 
\be
\omega^2 - (\Lambda \, d_i B_0 \kpa k) \, \omega - B_0^2 \kpa^2 = 0 \, ,
\label{dispersion}
\ee
with 
\be 
{{\tilde{\bf v}}_\kk \brace {\tilde{\bf b}}_\kk} = 
\Lambda \, i \, \ek \times {{\tilde{\bf v}}_\kk \brace {\tilde{\bf b}}_\kk} \, .
\ee
We obtain the solutions 
\be
\omega \equiv \ols = \frac{s \kpa k d_i B_0}{2} 
\left(s \Lambda + \sqrt{1 + \frac{4}{d_i^2 k^2}}\right) \, , 
\label{dispersion2}
\ee
where the value ($\pm 1$) of $s$ defines the directional wave polarity. In other words, 
we have $s \kpa \ge 0$ and $\ols$ is a positive frequency. 
The Alfv\'en wave polarization $\Lambda$ tells us if the wave is right ($\Lambda=s$) 
or left ($\Lambda=-s$) circularly polarized. In the first case, we are dealing with 
whistler waves, whereas in the latter case with ion cyclotron 
waves. We see that the transverse circularly polarized Alfv\'en waves are dispersive 
and we note that we recover the two well-known limits, \ie the pure whistler waves 
($\omega = s \kpa k d_i B_0$) in the high frequency limit ($k d_i \to \infty$), and 
the standard Alfv\'en waves ($\omega = s \kpa B_0$) in the low frequency limit 
($k d_i \to 0$). The Alfv\'en waves become linearly polarized only when the Hall term 
vanishes: when the Hall term is present, whatever its magnitude is, the Alfv\'en 
waves are circularly polarized. Note that this situation is different from the 
compressible case for which the Alfv\'en waves are elliptically polarized. As 
expected, it is possible to show (Hameiri et al., 2005; Sahraoui et al., 2005) that 
the ion cyclotron wave has a resonance at the frequency $\omega_{ci} \kpa/k$, 
where $\omega_{ci} = B_0 / d_i$. Therefore, with such an approximation, only whistler 
waves survive at high frequency.

\subsection{Complex helicity decomposition}
\label{chd}

Given the incompressibility constraints ($\ref{wave1}$) and ($\ref{wave4}$), it is 
convenient to project the Hall MHD equations in the plane orthogonal to $\kk$. We will 
use the complex helicity decomposition technique which has been shown to be effective 
in providing a compact description of the dynamics of 3D incompressible fluids (Craya, 
1958; Moffatt, 1970; Kraichnan, 1973; Cambon et al., 1989; Lesieur, 1990; Waleffe, 1992; 
Turner,2000; Galtier, 2003; Galtier and Bhattacharjee, 2003). The complex helicity basis 
is also particularly useful since it allows to diagonalize systems dealing with 
circularly polarized waves. We introduce the complex helicity decomposition
\be
{\bf h^{\Lambda}}(\kk) \equiv \hh = \ete + i \Lambda \efi \, ,
\label{basis0}
\ee
where 
\be
\ete = \efi \times \ek \, ,
\ee
\be
\efi = \frac{\ep \times \ek}{|\ep \times \ek|} \, ,
\label{efi}
\ee
and $|\ete (\kk)|$=$|\efi(\kk)|$=$1$. We note that ($\ek$, $h^+_{\kk}$, $h^-_{\kk}$) 
form a complex basis with the following properties:
\be
{\bf h^{-\Lambda}_{k}} = {\bf h^{\Lambda}_{-k}} \, ,
\ee
\be
\ek \times \hh = - i \Lambda \, \hh \, ,
\ee
\be
\kk \cdot \hh = 0 \, ,
\ee
\be
{\bf h^{\Lambda}_k} \cdot {\bf h^{\Lp}_k} = 2 \, \delta_{-\Lp \Lambda}\, .
\ee
We project the Fourier transform of the original vectors ${\bf v} ({\bf x})$ and 
${\bf b} ({\bf x})$ on the helicity basis:
\be
{\bf v}_\kk = \sum_{\Lambda} \, {\cal U}_{\Lambda} (\kk) \, {\bf h^{\Lambda}_k} 
= \sum_{\Lambda} \, {\cal U}_{\Lambda} \, {\bf h^{\Lambda}_k} \, ,
\ee
\be
{\bf b}_\kk = \sum_{\Lambda} \, {\cal B}_{\Lambda} (\kk) \, {\bf h^{\Lambda}_k} 
= \sum_{\Lambda} \, {\cal B}_{\Lambda} \, {\bf h^{\Lambda}_k} \, . 
\label{basis1}
\ee
We introduce expressions of the fields into the Hall MHD equations written in Fourier 
space and we multiply by vector ${\bf h^{\Lambda}_{-k}}$. First, we will focus on the 
linear dispersion relation ($\epsilon=0$) which reads:
\be
\partial_t {\cal Z}_{\Lambda}^s = - i \, \ols {\cal Z}_{\Lambda}^s \, ,
\label{zdispersion}
\ee
with 
\be
{\cal Z}_{\Lambda}^s \equiv {\cal U}_{\Lambda} + \xi_{\Lambda}^s {\cal B}_{\Lambda} \, ,
\label{zab}
\ee
\be
\xi_{\Lambda}^s (k) = \xi_{\Lambda}^s = - \frac{s d_i k}{2} 
\left( s \Lambda + \sqrt{1 + \frac{4}{d_i^2 k^2}} \right) \, . 
\ee
Equation ($\ref{zdispersion}$) shows that ${\cal Z}_{\Lambda}^s$ are the ``good'' variables 
for our system. These eigenvectors combine the velocity and the magnetic field in a non 
trivial way by a factor $\xi_{\Lambda}^s$ (with $\ols =-B_0 \kpa \xi_{\Lambda}^s$). 
In the large scale limit ($k d_i \to 0$), we see that $\xi_{\Lambda}^s \to - s$; we 
recover the Els\"asser variables used in standard MHD. In the small scale limit 
($k d_i \to \infty$), we have $\xi_{\Lambda}^s \to -s \, d_i k$, for $\Lambda = s$
(whistler waves), or $\xi_{\Lambda}^s \to (-s \, d_i k)^{-1}$, for $\Lambda = -s$. 
Therefore ${\cal Z}_{\Lambda}^s$ can be seen as a generalization of the Els\"asser 
variables to Hall MHD. In the rest of the paper, we will use the relation 
\be
{\cal Z}_{\Lambda}^s = (\alf - \alfm) \, \aak \, e^{-i \ols t} \, . 
\label{ZversusA}
\ee
where $\aak$ is the wave amplitude in the interaction representation for which we have, 
in the linear approximation, $\partial_t \aak = 0$. In particular, that means that weak 
nonlinearities will modify only slowly in time the Hall MHD wave amplitudes. The 
coefficient in front of the wave amplitude is introduced in advance to simplify the 
algebra that we are going to develop.

\section{Helical wave turbulence formalism}

\subsection{Fundamental equations}

We decompose the inviscid nonlinear Hall MHD equations (\ref{wave2})--(\ref{wave3}) on the 
complex helicity decomposition introduced in the previous section. Then we project the 
equations on vector ${\bf h^{\Lambda}_{-k}}$. We obtain:
\be
\partial_t {\cal U}_{\Lambda} - i B_0 \kpa {\cal B}_{\Lambda} = 
\ee
$$
- \frac{i \epsilon}{2} \int \sum_{\Lambda_p, \Lambda_q} 
({\cal U}_{\Lambda_p} {\cal U}_{\Lambda_q} - {\cal B}_{\Lambda_p} {\cal B}_{\Lambda_q})
(\kk \cdot {\bf h^{\Lambda_p}_p}) ({\bf h^{\Lambda_q}_q} \cdot {\bf h^{-\Lambda}_k})
\, \delta_{pq,k} \, d{\bf p} \, d{\bf q} \, , 
$$
and
\be
\partial_t {\cal B}_{\Lambda} - i B_0 \kpa {\cal U}_{\Lambda} + 
i \Lambda d_i B_0 \kpa k {\cal B}_{\Lambda} = 
\ee
$$
- \frac{i \epsilon}{2} \int \sum_{\Lambda_p, \Lambda_q}
({\cal U}_{\Lambda_p} {\cal B}_{\Lambda_q} - {\cal U}_{\Lambda_q} {\cal B}_{\Lambda_p})
(\kk \cdot {\bf h^{\Lambda_p}_p}) ({\bf h^{\Lambda_q}_q} \cdot {\bf h^{-\Lambda}_k})
\, \delta_{pq,k} \, d{\bf p} \, d{\bf q}
$$
$$
+ 
$$
$$
\frac{i \epsilon}{2} \int k \Lambda d_i \, \sum_{\Lambda_p, \Lambda_q}
{\cal B}_{\Lambda_p} {\cal B}_{\Lambda_q} 
[({\bf q} \cdot {\bf h^{-\Lambda}_k}) ({\bf h^{\Lambda_p}_p} \cdot {\bf h^{\Lambda_q}_q})
- ({\bf q} \cdot {\bf h^{\Lambda_p}_p}) ({\bf h^{\Lambda_q}_q} \cdot {\bf h^{-\Lambda}_k})]
\, \delta_{pq,k} \, d{\bf p} \, d{\bf q} \, ,
$$
where $\delta_{pq,k}=\delta({\bf p} + {\bf q} - \kk)$. The delta distributions come 
from the Fourier transforms of the nonlinear terms. We introduce the generalized 
Els\"asser variables $\aak$ in the interaction representation and we find: 
\be
\partial_t \aak = 
\frac{i \epsilon}{2} \int \sum_{\Lambda_p, \Lambda_q \atop s_p, s_q}
L{{\Lambda \Lambda_p \Lambda_q \atop s \, s_p \, s_q} \atop -k \, p \, q}
\, \aap \aaq \, e^{-i \Omega_{pq,k} t} \, \delta_{pq,k} \, d{\bf p} \, d{\bf q} \, ,
\label{fonda1}
\ee
where
\be
L{{\Lambda \Lambda_p \Lambda_q \atop s \, s_p \, s_q} \atop k \, \, p \, \, q} = 
\frac{1 - {\alf}^2}{\alf - \alfm} \, 
({\bf q} \cdot {\bf h^{\Lambda}_k}) ({\bf h^{\Lambda_p}_p} \cdot {\bf h^{\Lambda_q}_q})
\hspace{.5cm} +
\label{Lgeo}
\ee
$$
\frac{{\alf}^2 + \alf \alfmp - \alf \alfmq - \alfmp \alfmq}{\alf - \alfm} \, 
({\bf q} \cdot {\bf h^{\Lambda_p}_p}) ({\bf h^{\Lambda_q}_q} \cdot {\bf h^{\Lambda}_k}) \, ,
$$
and 
\be
\Omega_{pq,k} = \olsp + \olsq - \ols \, .
\ee
Equation (\ref{fonda1}) is the wave amplitude equation from which it is possible to 
extract some information. As expected we see that the nonlinear terms are of order 
$\epsilon$. This means that weak nonlinearities will modify only slowly in time the 
Hall MHD wave amplitude. They contain an exponentially oscillating term which is 
essential for the asymptotic closure. Indeed, wave turbulence deals with variations 
of spectral densities at very large time, \ie for a nonlinear transfer time much greater 
than the wave period. As a consequence, most of the nonlinear terms are destroyed by 
phase mixing and only a few of them, the resonance terms, survive (see \eg Newell et 
al., 2001). 
The expression obtained for the fundamental equation (\ref{fonda1}) is usual in wave 
turbulence. The main difference between problems is localized in the matrix $L$ 
which is interpreted as a complex geometrical coefficient. We will see below that the 
local decomposition allows to get a polar form for such a coefficient which is much 
easier to manipulate. 
From equation (\ref{fonda1}) we see eventually that, contrary to incompressible MHD, 
there is no exact solutions to the nonlinear problem in incompressible Hall MHD. The 
origin of such a difference is that in MHD the nonlinear term involves Alfv\'en waves 
traveling only in opposite directions whereas in Hall MHD this constrain does not 
exist (we have a summation over $\Lambda$ and $s$). In other words, if one type of 
wave is not present in incompressible MHD then the nonlinear term cancels whereas 
in incompressible Hall MHD it is not the case. The conclusion reached by Mahajan 
and Krishan (2005) is therefore not correct: the condition  
${\cal U}_{\Lambda} = - \xi_{\Lambda}^s {\cal B}_{\Lambda}$ does not correspond to 
a nonlinear solution of incompressible Hall MHD.

\subsection{Local decomposition}
\label{local}

\begin{figure}
\centerline{\psfig{figure=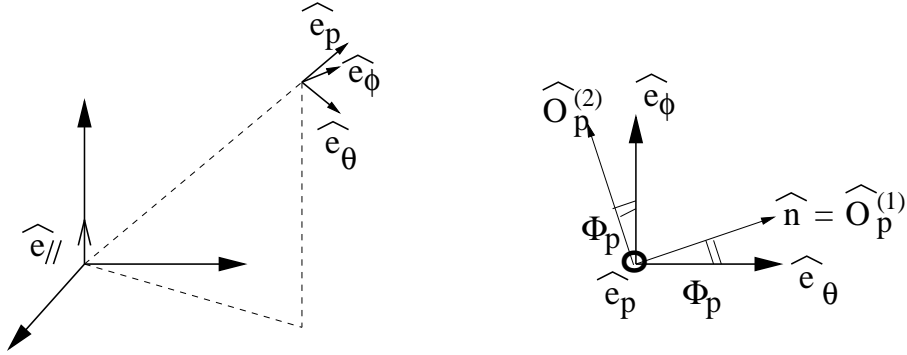,width=12cm,height=4.7cm}}
\caption{Sketch of the local decomposition for a given wavevector $\pp$.}
\label{localdecomp}
\end{figure}

In order to evaluate the scalar products of complex helical vectors found in the 
geometrical coefficient (\ref{Lgeo}), it is convenient to introduce a vector basis 
local to each particular triad (Waleffe, 1992; Turner, 2000; Galtier, 2003; Galtier 
and Bhattacharjee, 2003). For example, for a given vector $\pp$, we define the 
orthonormal basis vectors, 
\be
\begin{array}{lll}
{\bf {\hat O^{(1)}}}(\pp) &= \nn \, ,\\[.2cm]
{\bf {\hat O^{(2)}}}(\pp) &= \eep \times \nn \, ,\\[.2cm]
{\bf {\hat O^{(3)}}}(\pp) &= \eep \, ,
\end{array}
\label{basis2}
\ee
where $\eep=\pp/|\pp|$ and 
\be
\nn = {\pp \times \kk \over | \pp \times \kk |} =
{\qq \times \pp \over | \qq \times \pp |} =
{\kk \times \qq \over | \kk \times \qq |} \, .
\ee
We see that the vector $\nn$ is normal to any vector of the triad ($\kk$,$\pp$,$\qq$) 
and changes sign if $\pp$ and $\qq$ are interchanged, \ie 
$\nn_{(\kk,\qq,\pp)} = - \nn_{(\kk,\pp,\qq)}$. Note that $\nn$ does not change by cyclic 
permutation, \ie, $\nn_{(\kk,\qq,\pp)} = \nn_{(\qq,\pp,\kk)} =\nn_{(\pp,\kk,\qq)}$. 
A sketch of the local decomposition is given in Fig. \ref{localdecomp}. 
We now introduce the vectors
\be
\Xi^{\Lambda_p}(\pp) \equiv \Xi^{\Lambda_p}_\pp = 
{\bf {\hat O^{(1)}}}(\pp) + i \Lambda_p {\bf {\hat O^{(2)}}}(\pp) \, ,
\ee
and define the rotation angle $\Phi_p$, so that
\be
\begin{array}{lll}
\cos \Phi_p &= \nn \cdot \ete (\pp)\, , \\
\sin \Phi_p &= \nn \cdot \efi (\pp) \, .
\end{array}
\ee
The decomposition of the helicity vector ${\bf h^{\Lambda_p}_p}$ in the local basis
gives
\be
{\bf h^{\Lambda_p}_p} = \Xi^{\Lambda_p}_\pp \, e^{i \Lambda_p \Phi_p} \, . 
\ee
After some algebra we obtain the following polar form for the matrix $L$, 
\be
L{{\Lambda \Lambda_p \Lambda_q \atop s \, s_p \, s_q} \atop k \, \, p \, \, q} = 
\label{lint2}
\ee
$$
i \, e^{i (\Lambda \Phi_k + \Lambda_p \Phi_p + \Lambda_q \Phi_q)} \, 
\frac{\Lambda \, \Lambda_p \, \Lambda_q}{\alf - \alfm} \, \frac{\sin \psi_k}{k} \, 
[({\alf}^2 - \alfmp \alfmq) \, \Lambda_q q \, (\Lambda k + \Lambda_p p + \Lambda_q q) 
$$
$$
+
$$
$$
\alf \, (\alfmp  - \alfmq) \, k q \, (\Lambda \Lambda_q + \cos \psi_p)] \, .
$$
The angle $\psi_k$ refers to the angle opposite to $\kk$ in the 
triangle defined by $\kk=\pp+\qq$ 
($\sin \psi_k = \nn\cdot(\qq \times \pp)/|(\qq \times \pp)|$). 
To obtain equation (\ref{lint2}), we have also used the well-known
triangle relations 
\be
{\sin \psi_k \over k} = 
{\sin \psi_p \over p} = 
{\sin \psi_q \over q} \, . 
\ee
Further modifications have to be made before applying the spectral formalism. In particular, 
the fundamental equation has to be invariant under interchange of $\pp$ and $\qq$. To do 
so, we introduce the symmetrized and renormalized matrix $M$: 
\be
M{{\Lambda \Lambda_p \Lambda_q \atop s \, s_p \, s_q} \atop k \, \, p \, \, q} = 
i \, d_i \, \frac{1}{{\alf}^2} \, \frac{\alf - \alfm}{\alfq - \alfp} \, 
\left( L{{\Lambda \Lambda_p \Lambda_q \atop s \, s_p \, s_q} \atop k \, \, p \, \, q}  + 
L{{\Lambda \Lambda_q \Lambda_p \atop s \, s_q \, s_p} \atop k \, \, q \, \, p} \right) \, .
\ee
Finally, by using the identities given in Appendix \ref{relation}, we obtain:
\be
\partial_t \aak = 
\frac{\epsilon}{4 \, d_i} \int \sum_{\Lambda_p, \Lambda_q \atop s_p, s_q}
{\alf}^2 \, \frac{\alfq - \alfp}{\alf - \alfm} \, 
M{{\Lambda \Lambda_p \Lambda_q \atop s \, s_p \, s_q} \atop -k \, p \, q}
\, \aap \aaq \, e^{-i \Omega_{pq,k} t} \, \delta_{pq,k} \, d{\bf p} \, d{\bf q} \, ,
\label{fonda2}
\ee
where
\be
M{{\Lambda \Lambda_p \Lambda_q \atop s \, s_p \, s_q} \atop k \, \, p \, \, q} = 
\label{fonda2bis}
\ee
$$
e^{i (\Lambda \Phi_k + \Lambda_p \Phi_p + \Lambda_q \Phi_q)} \, 
\Lambda \, \Lambda_p \, \Lambda_q \, \frac{\sin \psi_k}{k} \, 
(\Lambda k + \Lambda_p p + \Lambda_q q) \, (1 - {\alfm}^2 {\alfmp}^2 {\alfmq}^2) \, .
$$
The matrix $M$ possesses the following properties ($*$ denotes the complex conjugate),
\be
\left(
M{{\Lambda \Lambda_p \Lambda_q \atop s \, s_p \, s_q} \atop k \, \, p \, \, q}
\right)^* = 
M{{-\Lambda -\Lambda_p -\Lambda_q \atop -s \, -s_p \, -s_q} \atop k \, \, \, p \, \, \, q} =
M{{\Lambda \, \, \, \, \Lambda_p \, \, \, \, \Lambda_q \atop s \, \, \, \, s_p 
\, \, \, \, s_q} \atop -k-p-q} \, , 
\label{prop1}
\ee
\be
M{{\Lambda \Lambda_p \Lambda_q \atop s \, s_p \, s_q} \atop k \, \, p \, \, q} = 
- M{{\Lambda \Lambda_q \Lambda_p \atop s \, s_q \, s_p} \atop k \, \, q \, \, p} \, ,
\label{prop2}
\ee
\be
M{{\Lambda \Lambda_p \Lambda_q \atop s \, s_p \, s_q} \atop k \, \, p \, \, q} = 
- M{{\Lambda_q \Lambda_p \Lambda \atop s_q \, s_p \, s} \atop q \, \, p \, \, k} \, ,
\label{prop3}
\ee
\be
M{{\Lambda \Lambda_p \Lambda_q \atop s \, s_p \, s_q} \atop k \, \, p \, \, q} = 
-M{{\Lambda_p \Lambda \Lambda_q \atop s_p \, s \, s_q} \atop p \, \, k \, \, q} \, .
\label{prop4}
\ee
Equation (\ref{fonda2}) is the fundamental equations that describe the slow evolution 
of the Alfv\'en wave amplitudes due to the nonlinear terms of the incompressible Hall 
MHD equations. It is the starting point for deriving the wave kinetic equations. 
The local decomposition used here allows us to represent concisely complex information 
in an exponential function. As we will see, it will simplify significantly the derivation 
of the wave kinetic equations. 

From equation (\ref{fonda2}) we note that the nonlinear coupling between helicity states 
associated with wavevectors, $\pp$ and $\qq$, vanishes when the wavevectors are collinear 
(since then, $\sin \psi_k=0$). This property is similar to the one found in the limit of 
EMHD (Galtier and Bhattacharjee, 2003). It seems to be a general property for helicity 
waves (Kraichnan,1973; Waleffe, 1993; Turner, 2000; Galtier, 2003). 
Additionally, we note that the nonlinear coupling between helicity states vanishes whenever 
the wavenumbers $p$ and $q$ are equal if their associated wave and directional polarities, 
$\Lambda_p$, $\Lambda_q$, and $s_p$, $s_q$ respectively, are also equal. In the case of
whistler (EMHD) waves, for which we have $\Lambda=s$ (right circularly polarized), this 
property was already observed (Galtier and Bhattacharjee, 2003). Here we generalize this 
finding to right and left circularly polarized waves. In the large scale limit, \ie when 
we tend to pure incompressible MHD, this property tends to disappear. For pure 
incompressible MHD, where Alfv\'en waves are linearly polarized, this is not observed 
anymore. As noticed before, the nature of the polarization seems to be fundamental. 

We are interested by the long-time behavior of the Alfv\'en wave amplitudes. From the 
fundamental equation (\ref{fonda2}), we see that the nonlinear wave coupling will come 
from resonant terms such that, 
\be
\left\{
\begin{array}{lll}
\kk = \pp + \qq \, , \\[.2cm]
\kpa \alf = p_{\parallel} \alfp + q_{\parallel} \alfq \, . 
\end{array}
\right.
\label{resonance0}
\ee
The resonance condition may also be written:
\be
\frac{\alf - \alfp}{q_{\parallel}} = 
\frac{\alfq - \alf}{p_{\parallel}} =
\frac{\alfq - \alfp}{k_{\parallel}} \, .
\label{resonance}
\ee
As we will see below, the relations (\ref{resonance}) are useful in simplifying the 
wave kinetic equations and demonstrating the conservation of ideal invariants. In 
particular, we note that we recover the resonance conditions for whistler waves by taking 
the appropriate limit.

\subsection{Dynamics and wave kinetic equations}

Fully developed wave turbulence is a state of a system composed of many simultaneously 
excited and interacting nonlinear waves where the energy distribution, far from 
thermodynamic equilibrium, is characterized by a wide power law spectrum. This range of 
wavenumbers, the inertial range, is generally localized between large scales at which 
energy is injected in the system and small dissipative scales. The origin of wave 
turbulence dates back to the early sixties and since then many papers have been devoted 
to the subject (see \eg Hasselmnan, 1962; Benney and Saffman, 1966; Zakharov, 1967; 
Benney and Newell, 1969; Sagdeev and Galeev, 1969; Kuznetsov, 1972; Zakharov et al.; 1992; 
Newell et al., 2001).
The essence of weak wave turbulence is the statistical study of large ensembles of 
weakly interacting dispersive waves via a systematic asymptotic expansion in powers of 
small nonlinearity. This technique leads finally to the derivation of wave kinetic 
equations for quantities like the energy and more generally for the quadratic invariants 
of the system studied. Here, we will follow the standard Eulerian formalism of wave 
turbulence (see \eg Benney and Newell, 1969). 

We define the density tensor $\qls(\kk)$ for an homogeneous turbulence, such that:
\be
\langle \aak(\kk) \, \aakp(\kkpr) \rangle \equiv 
\qls(\kk) \, \delta (\kk + \kkpr) \, \delta_{\Lambda \Lambda^{\prime}} \, 
\delta_{s s^{\prime}} \, , 
\label{tensor1}
\ee
for which we shall write a ``closure'' equation. The presence of the delta 
$\delta_{\Lambda \Lambda^{\prime}}$ and $\delta_{s s^{\prime}}$ means that correlations 
with opposite wave or directional polarities have no long-time influence in the wave 
turbulence regime; the third delta distribution $\delta (\kk + \kkpr)$ is the consequence 
of the homogeneity assumption. It is strongly linked to the form of the frequency 
$\ols$ (see the discussion in Section \ref{45}). Details of the derivation of the wave 
kinetic equations are given in Appendix \ref{derivation}. We obtain the following result: 
\be
\partial_t \qls (\kk) = 
\label{ke1}
\ee
$$
\frac{\pi \, \epsilon^2}{4 \, d_i^2 B_0^2} \int 
\sum_{\Lambda_p, \Lambda_q \atop s_p, s_q} 
\left( \frac{\sin \psi_k}{k} \right)^2 (\Lambda k + \Lambda_p p + \Lambda_q q)^2 
\left( 1 - {\alfm}^2 {\alfmp}^2 {\alfmq}^2 \right)^2 
\left( \frac{\alfq - \alfp}{\kpa} \right)^2 
$$
$$
\left( \frac{\ols}{1+{\alfm}^2} \right) 
\left[
\left( \frac{\ols}{1+{\alfm}^2} \right) \frac{1}{\qls(\kk)} - 
\left( \frac{\olsp}{1+{\alfmp}^2} \right) \frac{1}{\qlsp(\pp)} -
\left( \frac{\olsq}{1+{\alfmq}^2} \right) \frac{1}{\qlsq(\qq)}
\right]
$$
$$
\qls (\kk) \, \qlsp (\pp) \, \qlsq (\qq) \, 
\delta(\Omega_{k,pq})\, \delta_{k,pq} \, d{\bf p} \, d{\bf q} \, . 
$$
Equation (\ref{ke1}) is the main result of the helical wave turbulence formalism. 
It describes the statistical properties of incompressible Hall MHD wave turbulence 
at the lowest order, \ie for three-wave interactions.

\section{General properties of Hall MHD wave turbulence}

\subsection{Triadic conservation of ideal invariants}

In section \ref{invariants}, we have introduced the 3D ideal invariants of 
incompressible Hall MHD. The first test that the wave kinetic equations have to 
satisfy is the detailed conservation of these invariants, that is to say, the 
conservation of invariants for each triad ($\kk$, $\pp$, $\qq$). 
Starting from definitions (\ref{I1})--(\ref{I3}), we note that the Fourier spectra of 
the ideal invariants are: 
\be
E(\kk) = \sum_{\Lambda, s} (1 + {\alfm}^2) \, \qls (\kk) \, ,
\ee
\be
H_m(\kk) = \sum_{\Lambda, s} \frac{\Lambda}{k} \, \qls (\kk) \, ,
\ee
\be
H_G(\kk) = \sum_{\Lambda, s} \frac{\Lambda {\alfm}^4}{k} \, \qls (\kk) \, . 
\ee

We will first check the energy conservation. From expression (\ref{ke1}), we find:
\be
\partial_t E (t) \equiv 
\partial_t \int E (\kk) \, d\kk \equiv 
\partial_t  \int \sum_{\Lambda, s} e_{\Lambda}^s (\kk) \, d\kk = 
\label{kenergy0}
\ee
$$
\frac{\pi \, \epsilon^2}{4 \, d_i^2 B_0^2} \int 
\sum_{\Lambda, \Lambda_p, \Lambda_q \atop s, s_p, s_q} 
\left( \frac{\sin \psi_k}{k} \right)^2 (\Lambda k + \Lambda_p p + \Lambda_q q)^2 
\left( 1 - {\alfm}^2 {\alfmp}^2 {\alfmq}^2 \right)^2 
\left( \frac{\alfq - \alfp}{\kpa} \right)^2 
$$
$$
\ols \, 
\left[ \frac{\ols}{e_{\Lambda}^s(\kk)} - 
\frac{\olsp}{e_{\Lambda_p}^{s_p}(\pp)} -
\frac{\olsq}{e_{\Lambda_q}^{s_q}(\qq)} \right]
\qls (\kk) \, \qlsp (\pp) \, \qlsq (\qq) \, 
\delta(\Omega_{k,pq})\, \delta_{k,pq} \, d\kk \, d\pp \, d\qq \, . 
$$
Equation (\ref{kenergy0}) is invariant under cyclic permutations of wavevectors. 
That leads to: 
\be
\partial_t E (t) = 
\label{kenergy1}
\ee
$$
\frac{\pi \, \epsilon^2}{12 \, d_i^2 B_0^2} \int 
\sum_{\Lambda, \Lambda_p, \Lambda_q \atop s, s_p, s_q} 
\left( \frac{\sin \psi_k}{k} \right)^2 (\Lambda k + \Lambda_p p + \Lambda_q q)^2 
\left( 1 - {\alfm}^2 {\alfmp}^2 {\alfmq}^2 \right)^2 
\left( \frac{\alfq - \alfp}{\kpa} \right)^2 
$$
$$
\Omega_{k,pq} \, 
\left[ \frac{\ols}{e_{\Lambda}^s(\kk)} - 
\frac{\olsp}{e_{\Lambda_p}^{s_p}(\pp)} -
\frac{\olsq}{e_{\Lambda_q}^{s_q}(\qq)} \right]
\qls (\kk) \, \qlsp (\pp) \, \qlsq (\qq) \, 
\delta(\Omega_{k,pq})\, \delta_{k,pq} \, d\kk \, d\pp \, d\qq \, . 
$$
Total energy is conserved exactly on the resonant manifold since then $\Omega_{k,pq}=0$:
we have triadic conservation of total energy. 

For the other invariants, it is straightforward to show that the difference between 
them is conserved by the wave kinetic equations. With relation (\ref{ident5}), we 
obtain: 
\be
\partial_t (H_m (t) - H_G (t)) \equiv 
\partial_t \int (H_m (\kk) - H_G (\kk)) \, d\kk = 
\label{khelicity1}
\ee
$$
\frac{\pi \, \epsilon^2}{4 \, d_i B_0^2} \int 
\sum_{\Lambda, \Lambda_p, \Lambda_q \atop s, s_p, s_q} 
\left( \frac{\sin \psi_k}{k} \right)^2 (\Lambda k + \Lambda_p p + \Lambda_q q)^2 
\left( 1 - {\alfm}^2 {\alfmp}^2 {\alfmq}^2 \right)^2 
\left( \frac{\alfq - \alfp}{\kpa} \right)^2 
$$
$$
\kpa \, 
\left[
\left( \frac{\ols}{1+{\alfm}^2} \right) \frac{1}{\qls(\kk)} - 
\left( \frac{\olsp}{1+{\alfmp}^2} \right) \frac{1}{\qlsp(\pp)} -
\left( \frac{\olsq}{1+{\alfmq}^2} \right) \frac{1}{\qlsq(\qq)}
\right]
$$
$$
\qls (\kk) \, \qlsp (\pp) \, \qlsq (\qq) \, 
\delta(\Omega_{k,pq})\, \delta_{k,pq}  \, d\kk \, d\pp \, d\qq \, . 
$$
Equation (\ref{kenergy0}) is also invariant under cyclic permutations of wavevectors. 
Then one is led to 
\be
\partial_t (H_m (t) - H_G (t)) = 
\label{khelicity2}
\ee
$$
\frac{\pi \, \epsilon^2}{12 \, d_i B_0^2} \int 
\sum_{\Lambda, \Lambda_p, \Lambda_q \atop s, s_p, s_q} 
\left( \frac{\sin \psi_k}{k} \right)^2 (\Lambda k + \Lambda_p p + \Lambda_q q)^2 
\left( 1 - {\alfm}^2 {\alfmp}^2 {\alfmq}^2 \right)^2 
\left( \frac{\alfq - \alfp}{\kpa} \right)^2 
$$
$$
(\kpa - p_{\pa} - q_{\pa}) \, 
\left[
\left( \frac{\ols}{1+{\alfm}^2} \right) \frac{1}{\qls(\kk)} - 
\left( \frac{\olsp}{1+{\alfmp}^2} \right) \frac{1}{\qlsp(\pp)} -
\left( \frac{\olsq}{1+{\alfmq}^2} \right) \frac{1}{\qlsq(\qq)}
\right]
$$
$$
\qls (\kk) \, \qlsp (\pp) \, \qlsq (\qq) \, 
\delta(\Omega_{k,pq})\, \delta_{k,pq}  \, d\kk \, d\pp \, d\qq \, ,
$$
which is exactly equal to zero on the resonant manifold: we have triadic conservation 
of the difference between magnetic and global helicity. The last sufficient step would 
be to show detailed conservation for one of the two helicities. Unfortunately it is not 
so trivial and we will not give the proof here. However, we note already that magnetic 
helicity is conserved for equilateral triangles ($k=p=q$), \ie for strongly local 
interactions.

\subsection{General properties}

From the wave kinetic equations (\ref{ke1}), we find several general properties. 
Some of them can be obtained directly from the wave amplitude equation (\ref{fonda2}) 
as explained in Section \ref{local}. 
First, we observe that there is no coupling between Hall MHD waves associated with 
wavevectors, $\pp$ and $\qq$, when the wavevectors are collinear ($\sin \psi_k=0$). 
Second, we note that there is no coupling between helical waves associate with these 
vectors whenever the magnitudes, $p$ and $q$, are equal if their associated polarities, 
$s_p$ and $s_q$ in one hand and, $\Lambda_p$ and $\Lambda_q$ on the other hand, are 
also equal (since then, $\alfq - \alfp=0$). 
These properties hold for the three inviscid invariants and generalize what 
was found previously in EMHD (Galtier and Bhattacharjee, 2003) where we only have 
right circularly polarized waves ($\Lambda=s$). It seems to be a generic property of 
helical wave interactions (Kraichnan,1973; Waleffe, 1993; Turner, 2000; Galtier, 2003). 
As noted before, this property tends to disappear when the large scale limit is taken, 
\ie when we tend to standard MHD. 
Third, it follows from the previous observations that a strong helical perturbation 
localized initially in a narrow band of wavenumbers will lead to a weak transfer of 
energy, magnetic and global helicities. Note that these properties can be inferred 
from the fundamental equation (\ref{fonda2}) as well.

\subsection{High frequency limit of electron MHD}

In the present section we shall demonstrate that the small scale limit 
($d_ik \to +\infty$) of the wave kinetic equations (\ref{ke1}) tends to the 
expected equations of electron MHD when only right ($\Lambda=s$) circularly 
polarized waves are taken into account. We recall that, in the small scale limit, 
$\xi_{s}^{s} \to -sd_ik$ and $\xi_s^{-s} \to s/d_ik$. One obtains
\be
\partial_t q^s_s (\kk) = 
\label{ke1_emhdlimit}
\ee
$$
\frac{\pi \, d_i^2 \, \epsilon^2}{4} \int 
\sum_{s_p, s_q} 
\left( \frac{\sin \psi_k}{k} \right)^2 (sk + s_p p + s_q q)^2 
\left( \frac{s_q q - s_p p}{\kpa} \right)^2 s k \kpa 
$$
$$
\left[ s k \kpa \, q_{s_p}^{s_p} (\pp) \, q_{s_q}^{s_q} (\qq) - 
s_p p p_{\parallel} \, q^s_s (\kk) \, q_{s_q}^{s_q} (\qq) - 
s_q q q_{\parallel} \, q^s_s (\kk) \, q_{s_p}^{s_p} (\pp) \right]
\delta(\Omega_{k,pq})\, \delta_{k,pq} \, d{\bf p} \, d{\bf q} \, , 
$$
where $\Omega_{k,pq}= B_0\,d_i(s\kpa k - s_p p_{\parallel}p - s_q q_{\parallel} q)$. 
The kinetic equations found have exactly the same form as in Galtier and Bhattacharjee 
(2003) (where by definition $\omega_s^s= s \, \omega_k$, $B_0=1$ and $d_i=1$) who 
derived a wave turbulence theory for electron MHD (see the discussion in Section 
\ref{mastersequa}). In particular, this means 
that by recovering the electron MHD description from the present Hall MHD theory, 
we recover all the properties already found for whistler wave turbulence (anisotropy, 
scaling laws, direct cascade...). We will come back to that point in Section 
\ref{mastersequa}. Note finally that there is a strong analogy between such a limit 
and wave turbulence in rapidly rotating flows (Galtier, 2003; Bellet et al., 2005; 
Morize et al., 2005). The physical reason is that in both problems (i) there is a 
privileged direction, played by the rotating axis or the magnetic field $\bB_0$, 
(ii) there are dispersive helical waves, called inertial waves for rotating flows, 
and (iii) the wave frequencies are not very different, \ie the inertial wave 
frequency is proportional to $\kpa / k$. In the past, comparisons have been made 
between incompressible rotating turbulence and magnetized plasmas described by 
incompressible MHD. The results obtained in the framework of wave turbulence show 
clearly that the comparison is much more relevant if one considers electron MHD 
plasmas. Indeed, a small nonlinear transfer is found along the privileged direction 
for rotating and electron MHD turbulence whereas this kind of transfer is strictly 
forbidden in MHD turbulence.

\subsection{High frequency limit of ion MHD}

We may also be interested by the small scale limit ($d_ik \to +\infty$) of the wave 
kinetic equations (\ref{ke1}) when only left ($\Lambda=-s$) circularly polarized 
waves are taken into account. We decide to call this limit the ion MHD (IMHD) 
approximation following the example of the electron MHD. We have, for such a limit, 
$\xi_{-s}^{-s} \to sd_ik$ and $\xi_{-s}^{s} \to -s/d_ik$. One finds 
\be
\partial_t q_{-s}^s (\kk) = 
\label{ke1_leftlimit}
\ee
$$
\frac{\pi \, d_i^2 \epsilon^2}{4} \int 
\sum_{s_p, s_q} 
\left( \frac{\sin \psi_k}{k} \right)^2 (s k + s_p p + s_q q)^2 \, 
\left( \frac{s_q q - s_p p}{pq\kpa} \right)^2 s\kpa k \, p^4 q^4 \, 
\delta(\Omega_{k,pq}) \, \delta_{k,pq}
$$
$$
\left[
\frac{s\kpa}{k^3} \, q_{-s_p}^{s_p} (\pp) \, q_{-s_q}^{s_q} (\qq) - 
\frac{s_p p_{\parallel}}{p^3} \, q_{-s}^s (\kk) \, q_{-s_q}^{s_q} (\qq) -
\frac{s_q q_{\parallel}}{q^3} \, q_{-s}^s (\kk) \, q_{-s_p}^{s_p}(\pp)
\right] d{\bf p} \, d{\bf q} \, , 
$$
with $\Omega_{k,pq}= (B_0/d_i)(s\kpa/k - s_p p_{\parallel}/p - s_q q_{\parallel}/q)$.
No existing theory has been developed for such a limit. We will see in Section 
\ref{mastersequa} that it is possible to extract from the master equations of 
ion cyclotron turbulence some exact properties.

\subsection{Low frequency limit of standard MHD}
\label{45}

This section is devoted to the low frequency limit, \ie the large scale limit, of 
the wave kinetic equations (\ref{ke1}) of Hall MHD. The MHD limit is somewhat 
singular for Hall MHD. Indeed the large scale limit does not tend to the expected 
wave kinetic equations that were derived first by Galtier et al. (2000). The subtle 
point resides in the kinematics: for pseudo-dispersive MHD waves, the definition 
(\ref{tensor1}) of the density tensor $\qls(\kk)$ that is used for Hall MHD is not 
valid anymore. The reason is that in the large scale limit the polarization
$\Lambda$ does not appear anymore in the frequency, which is $\omega=s \kpa B_0$. 
Then the kinematics tells us that the definition for the density tensor is 
\be
\langle \aak(\kk) \, \aakp(\kkpr) \rangle \equiv 
{\tilde q}_{\Lambda \Lambda^{\prime}}^{s s^{\prime}} \, 
\delta (\kk + \kkpr) \, \delta_{s s^{\prime}} \, , 
\label{tensor2}
\ee
where the condition $\Lambda=\Lambda^{\prime}$ is not necessary satisfied. This 
means that the large scale limit of equations (\ref{ke1}) leads to MHD wave 
kinetic equations in the particular case where helicity terms are supposed to be 
absent and where equality between shear-Alfv\'en and pseudo-Alfv\'en waves energy 
is assumed. Indeed, helicity terms involve quantities for which 
$\Lambda=-\Lambda^{\prime}$ and the total energy involves only terms for which 
$\Lambda=\Lambda^{\prime}$. However, energies for shear-Alfv\'en waves and 
pseudo-Alfv\'en waves involve terms with different polarities. 
For shear-Alfv\'en waves, we have:
\be
{\bf v}_\kk - s {\bf b}_\kk = 
{\cal Z}_+^s {\bf h^+_k} + {\cal Z}_-^s {\bf h^-_k} =
({\cal Z}_+^s - {\cal Z}_-^s ) \, i \, \efi \, ,
\ee
and for pseudo-Alfv\'en waves:
\be
{\bf v}_\kk - s {\bf b}_\kk = 
{\cal Z}_+^s {\bf h^+_k} + {\cal Z}_-^s {\bf h^-_k} =
({\cal Z}_+^s + {\cal Z}_-^s ) \, \ete \, . 
\ee
Energies associated to shear- and pseudo-Alfv\'en waves are respectively:
\be
({\cal Z}_+^s - {\cal Z}_-^s )({\cal Z}_+^s - {\cal Z}_-^s )^*=
|{\cal Z}_+^s|^2 + |{\cal Z}_-^s|^2 - {\cal Z}_+^s ({\cal Z}_-^s)^*
- {\cal Z}_-^s ({\cal Z}_+^s)^* \, ,
\ee
and
\be
({\cal Z}_+^s + {\cal Z}_-^s )({\cal Z}_+^s + {\cal Z}_-^s )^*=
|{\cal Z}_+^s|^2 + |{\cal Z}_-^s|^2 + {\cal Z}_+^s ({\cal Z}_-^s)^*
+ {\cal Z}_-^s ({\cal Z}_+^s)^* \, . 
\ee
We see clearly that if, in the wave kinetic equations, we only take into account 
(quadratic) 
terms with the same polarizations ($\Lambda$) then it is equivalent to assume equality 
between shear- and pseudo-Alfv\'en wave energies. We will see below that the large 
scale limit of the wave kinetic equations (\ref{ke1}) of Hall MHD tends to the 
expected MHD counterpart when the previous assumptions about helicities and energies 
are satisfied. A derivation of the wave kinetic equations is given in Appendix 
\ref{AppenD}. The result is given for the density tensor 
$q_{\Lambda \Lambda^{\prime}}^{s s^{\prime}}$. Contrary to Hall MHD, and actually 
to any problem in wave turbulence, principal value terms appear for incompressible 
MHD. The reason of the presence of principal value terms is linked to the nature 
of Alfv\'en waves which are pseudo-dispersive. 

Further comparisons between the results in Appendix \ref{AppenD} and the wave
kinetic equations obtained by Galtier et al. (2000) are of course possible 
(see \eg the discussion in Section \ref{mastersequa}) but, 
for simplicity, we prefer to focus our attention to the case where the density 
tensor is symmetric in $\Lambda$. Therefore we start our analysis with the 
general kinetic equation (\ref{ke1}) and take the large scale limit ($d_ik \to 0$) 
for which we have, at the leading order, $\alf \to -s - \Lambda d_i k/2$. 
After some simplifications, we arrive at:
\be
\partial_t \qls (\kk) = 
\label{ke_alfven}
\ee
$$
\frac{\pi \, \epsilon^2}{16} \int 
\sum_{\Lambda_p, \Lambda_q \atop s_p, s_q} 
\left( \frac{\sin \psi_k}{k} \right)^2 
(\Lambda k + \Lambda_p p + \Lambda_q q)^2 
\left( sk\Lambda + s_pp\Lambda_p + s_qq\Lambda_q \right)^2 
\left( \frac{s_q - s_p}{\kpa} \right)^2 
$$
$$
s \kpa \, \left(
s\kpa \qlsp(\pp) \qlsq(\qq) - s_pp_{\parallel} \qls(\kk) \qlsq(\qq) - 
s_qq_{\parallel} \qls(\kk) \qlsp(\pp) 
\right) \, 
\delta(\Omega_{k,pq})\, \delta_{k,pq} \, d{\bf p} \, d{\bf q} \, , 
$$
where $\Omega_{k,pq}=B_0(s\kpa - s_p p_{\parallel} - s_q q_{\parallel})$.
Note that we only have a nonlinear contribution when the wave polarities $s_p$ 
and $s_q$ are different. We recover here a well-known property of incompressible 
MHD: in such a limit, we only have nonlinear interactions between Alfv\'en waves 
propagating in different directions. One expands the summation over the directional 
polarities $s_p$ and $s_q$, and finds
\be
\partial_t \qls (\kk) = 
\label{ke_alfven2}
\ee
$$
\frac{\pi \, \epsilon^2}{4 B_0} \int 
\sum_{\Lambda_p, \Lambda_q} 
\left( \frac{\sin \psi_k}{k} \right)^2 \, 
(\Lambda k + \Lambda_p p + \Lambda_q q)^2 \, 
(\Lambda k + \Lambda_p p - \Lambda_q q)^2 
$$
$$
q_{\Lambda_q}^{-s} (\qq) \left[ q_{\Lambda_p}^s (\pp)- \qls (\kk) \right]
\, \delta(q_{\parallel}) \, \delta_{k,pq} \, d{\bf p} \, d{\bf q} \, . 
$$
This result is exactly the same as in Appendix \ref{AppenE} when MHD wave 
kinetic equations are considered in the particular case where only terms 
symmetric in $\Lambda$ are retained, \ie terms like 
${\tilde q}_{\Lambda \Lambda}^{ss}$. Therefore under these assumptions the MHD 
description does not appear like a singular limit for Hall MHD. 
(Note that this continuity in the description is {\it a priori} not so trivial 
if we remember that as soon as the Hall term is included in the standard MHD 
equations, whatever its magnitude is, it changes the polarity of the Alfv\'en 
waves which become circularly polarized.) 

The comparison with the wave kinetic equations derived by Galtier et al. (2000) 
(see equations (26)) is not direct since here the problem has been decomposed at 
the origin on a complex helicity basis. The main signature of this decomposition 
is the dependency on the wave polarity $\Lambda$. 
In spite of this difficulty, a common point is clearly seen in the 
presence of the delta $\delta(q_{\parallel})$ which arises because of the three-wave 
frequency resonance condition. This means that in any triadic resonant interaction, 
there is always one wave that corresponds to a purely 2D motion ($q_{\parallel}=0$)  
whereas the two others have equal parallel components ($p_{\parallel}=\kpa$). 
The direct consequence is the absence of nonlinear transfer along ${\bf B_0}$, 
a result predicted earlier by several authors (see \eg Montgomery and Turner, 1981; 
Shebalin et al., 1983). In other words, we have a two-dimensionalization of the 
Alfv\'en wave turbulence (see also \eg Ng and Bhattacharjee, 1997; Lithwick and 
Goldreich, 2003).

\section{Master equations of Hall MHD turbulence}
\label{mastersequa}
\subsection{General case}
\label{mastersequa2}

In order to extract further information about Hall MHD turbulence, we are going 
to write the expression of the spectral density $\qls (\kk)$ in terms of the 
invariants. In practice, we need to add a fourth variable, $E_d$, which has been 
chosen to be the difference between the kinetic and magnetic energy, namely
\be
E_d(\kk) = \sum_{\Lambda, s} ({\alfm}^2 -1) \, \qls (\kk) \, . 
\ee
Note that contrary to the MHD case, wave turbulence in Hall MHD allows to have 
a departure from equipartition between kinetic and magnetic energy and therefore 
a non trivial value for $E_d$. We will see, in Section \ref{nonlocality}, that this
property is fundamental to get non trivial nonlocal interactions between waves.
It is possible to inverse the system $\qls (E, E_d, Hm, H_G)$; one finds 
\be
\qls (\kk) = 
\label{qexpand}
\ee
$$
\frac{ ({\alf}^2+{\alfm}^2) [ ({\alf}^2-1) E(\kk) - ({\alf}^2+1) E_d(\kk) ] +
2 \Lambda k ({\alf}^4 H_m(\kk) - H_G(\kk) )}{4 \, ({\alf}^4-{\alfm}^4)} \, . 
$$
The introduction of expression (\ref{qexpand}) into (\ref{ke1}) leads to the wave 
kinetic equations for $E$, $E_d$, $Hm$ and $H_G$. We will not make such a lengthy 
development and we will rather focus on the master equations which drive the Hall 
MHD turbulence. Indeed, because of the presence of the factor $\Lambda$ in 
expression (\ref{qexpand}), we see that the nonlinear terms with different 
polarities will not play the same role. In particular, for the wave kinetic 
equations of energies, only the interactions between nonlinear terms involving either
energies or helicities will give a contribution. For the wave kinetic equations 
of helicities, we only have contributions from nonlinear terms involving 
energies and helicities. Thus the energy equations are the master equations 
driving Hall MHD turbulence. In other words, this means that an initial state 
with zero helicity will not generate any helicity at any scale. However, an 
initial state of zero helicity does not preclude the development of energy 
spectra. Assuming $H_m(\kk)=0$ and $H_G(\kk)=0$, the master kinetic equations of 
Hall MHD turbulence, written for the kinetic energy $E^V$ and the magnetic energy 
$E^B$, are 
\be
\partial_t {E^V (\kk) \brace E^B(\kk)} = 
\label{mastersHall}
\ee
$$
\frac{\pi \, \epsilon^2}{8 \, d_i^2 B_0^2} \int 
\sum_{\Lambda, \Lambda_p, \Lambda_q \atop s, s_p, s_q} 
\left( \frac{\sin \psi_k}{k} \right)^2 \, 
\frac{(\Lambda k + \Lambda_p p + \Lambda_q q)^2 
\, \left( 1 - {\alfm}^2 {\alfmp}^2 {\alfmq}^2 \right)^2}
{(1 + {\alfm}^2)(1 + {\alfmp}^2)(1 + {\alfmq}^2)}
$$
$$
\left( \frac{\alfq - \alfp}{\kpa} \right)^2 \, 
{{\alfm}^2 \brace 1} \, \frac{\ols \, \olsp}{{\alfm}^2 +1}
\left(
\frac{{\alfmq}^2 E^V(\qq) - E^B(\qq)}{{\alfmq}^2-1}
\right)
$$
$$
\left[ 
\left( \frac{{\alfmp}^2 E^V(\pp) - E^B(\pp)}{{\alfmp}^2-1} \right) -
\left( \frac{{\alfm}^2 E^V(\kk) - E^B(\kk)}{{\alfm}^2-1} \right)
\right] 
\delta(\Omega_{k,pq})\, \delta_{k,pq} \, d{\bf p} \, d{\bf q} \, . 
$$
In Appendix \ref{AppenC} the equivalent kinetic equations for the 
variables $E$ and $E_d$ are given. Below we describe the small scale 
(whistler and ion cyclotron waves) and large scale (pure Alfv\'en waves) 
limits of such a description. In particular, we will see that the role 
played by the kinetic and magnetic energies may be very different and 
that the master equations can be simplified further.

\subsection{Master equations in the limit of Whistler wave turbulence}

In the small scale limit ($d_ik \to +\infty$), one can distinguish between 
the whistler branch and the ion cyclotron branch. For right circularly 
polarized wave ($\Lambda=s$) we have ${\alfm}^2 \to 1/d_i^2k^2$, whereas for 
left circularly polarized wave ($\Lambda=-s$), ${\alfm}^2 \to d_i^2k^2$. 
Then in the limit of whistler wave turbulence (EMHD limit), we obtain
\be
\partial_t {E^V (\kk) \brace E^B(\kk)} = 
\label{mastersEMHD}
\ee
$$
\frac{\pi \, \epsilon^2 \, d_i^2}{8} \int 
\sum_{s, s_p, s_q} 
\left( \frac{\sin \psi_k}{k} \right)^2 \, 
(s k + s_p p + s_q q)^2 \, 
\left( \frac{s_q q - s_p p}{\kpa} \right)^2 
$$
$$
{1/d_i^2k^2 \brace 1} \, s \kpa k s_p p_{\parallel} p \, 
E^B(\qq) \left[ E^B(\pp) - E^B(\kk) \right] 
\delta(\Omega_{k,pq})\, \delta_{k,pq} \, d{\bf p} \, d{\bf q} \, , 
$$
with 
$\Omega_{k,pq}= B_0\,d_i(s\kpa k - s_p p_{\parallel}p - s_q q_{\parallel} q)$. 
We see clearly that the kinetic energy is not a relevant quantity in such a limit 
since the factor $1/d_i^2k^2$ damps the nonlinear contributions. This is not in 
contradiction with what we know if we remember that the velocity in the original 
Hall MHD equations is a combination of the ion and electron velocities. In the EMHD 
limit, \ie in the small spatio-temporal scale limit, this velocity tends to be 
small: ions do not have time to follow electrons and they provide a static 
homogeneous background on which electrons move. Since the magnetic field is frozen 
into the ideal electron flow (see Section \ref{invariants} for the proof) we understand 
why the magnetic energy is the dominant contribution to the wave kinetic equations. 
The wave kinetic equation found for the magnetic energy have, at leading order, 
exactly the same form as in Galtier and Bhattacharjee (2003) 
(where $\omega_s^s= s \, \omega_k$ and $B_0=1$) but they are more general in the sense 
that we have at our disposal not only an equation for the magnetic field, the dominant 
contribution, but also for the velocity that behaves like the magnetic field. The 
consequence is that we recover all the properties already known. In particular, in 
such a regime the energy spectrum follows the power law 
$E(\kpn,\kpa) \sim \kpn^{-5/2} \kpa^{-1/2}$ ($\kpa$ is assumed positive) which is an 
exact solution of the wave kinetic equations.

\subsection{Master equations in the limit of ion cyclotron wave turbulence}

For the ion cyclotron branch ($\Lambda=-s$), the small scale limit 
($d_ik \to +\infty$) gives
\be
\partial_t {E^V (\kk) \brace E^B(\kk)} = 
\label{mastersIon}
\ee
$$
\frac{\pi \, \epsilon^2}{8 \, d_i^2} \int 
\sum_{s, s_p, s_q} 
\left( \frac{\sin \psi_k}{k} \right)^2 \, 
(s k + s_p p + s_q q)^2 \, \left( \frac{s_p p - s_q q}{\kpa \, p \, q} \right)^2
$$
$$
{d_i^2 k^2 \brace 1} \, 
\frac{s \kpa s_p p_{\parallel} p \, q^2}{k} \, 
E^V(\qq) 
\left[ E^V(\pp) - E^V(\kk) \right] 
\delta(\Omega_{k,pq})\, \delta_{k,pq} \, d{\bf p} \, d{\bf q} \, , 
$$
with $\Omega_{k,pq}= (B_0/d_i)(s\kpa/k - s_p p_{\parallel}/p - s_q q_{\parallel}/q)$.
In such a limit, we are able to describe both the kinetic and the magnetic energies 
but we see that the dominant quantity is the kinetic energy since the factor 
$d_i^2k^2$ amplifies the nonlinear contributions in the corresponding equation 
(see \eg Mahajan and Krishan, 2005). The physical reason is that in such a limit, 
we are mainly dealing with ions whereas the magnetic field is frozen into the 
ideal electron flow (see Section \ref{invariants}). 
Like for the EMHD limit, it is possible to derive the exact power law solutions. 
First of all, we note that for strongly local interactions the resonance condition 
writes
\be
\left(\frac{s_p-s_q}{\kpa}\right)^2 \approx \left(\frac{s_q-s}{p_{\pa}}\right)^2 
\approx \left(\frac{s-s_p}{q_{\pa}}\right)^2 \, ,
\label{reson2}
\ee
which leads, as for the whistler waves case, to anisotropic turbulence. The same 
analysis for strongly nonlocal interactions leads to the same conclusion. Thus, 
we may write the wave kinetic equations in the limit $\kpn \gg \kpa$. For an 
axisymmetric turbulence $E^V({\bf \kpn},\kpa) = E^V(\kpn,\kpa) / 2\pi \kpn$; we 
find
\be
\partial_t E^V (\kpn,\kpa) = 
\label{mastersIon2}
\ee
$$
\frac{\epsilon^2}{16} \int 
\sum_{s, s_p, s_q} \sin \psi_{q_{\perp}} \, 
(s k_{\perp} + s_p p_{\perp} + s_q q_{\perp})^2 \, 
\left( \frac{s_p p_{\perp} - s_q q_{\perp}}{\kpa \, p_{\perp} \, q_{\perp}} \right)^2
s \kpa s_p p_{\parallel}
$$
$$
E^V(q_{\perp},q_{\pa}) 
\left[ \kpn E^V(p_{\perp},p_{\pa}) - p_{\perp} E^V(\kpn,\kpa) \right] 
\delta(\Omega_{k,pq})\, \delta_{\kpa, p_{\pa} q_{\pa}} \, 
dp_{\perp} dq_{\perp} dp_{\pa} dq_{\pa} \, , 
$$
where $\Omega_{k,pq}=(B_0/d_i)(s\kpa/\kpn - s_p p_{\parallel}/p_{\perp} - 
s_q q_{\parallel}/q_{\perp})$. Note that in such a limit the well-known
triangle relations are: 
$\sin \psi_{\kpn} / \kpn = \sin \psi_{p_{\perp}} / p_{\perp} = 
\sin \psi_{q_{\perp}} / q_{\perp}$. 
The wave kinetic equation (\ref{mastersIon2}) is now symmetric enough to apply 
a conformal transformation, called the Kuznetsov-Zakharov transformation. This 
transformation, a two-dimensional generalization of the Zakharov transformation, 
has been applied in various anisotropic problems (Kuznetsov, 1972; 
Balk et al. , 1990; Kuznetsov, 2001; Galtier and Bhattacharjee, 2003; Galtier, 2003). 
The bihomogeneity of the collision integrals in the wavenumbers $\kpn$ and $\kpa$ 
allows us to use the transformation
\be
\begin{array}{lll}
p_{\perp} &\to & \kpn^2 / p_{\perp} \, , \\[.2cm]
q_{\perp} &\to & \kpn q_{\perp} / p_{\perp} \, , \\[.2cm]
p_{\pa} &\to & \kpa^2 / p_{\pa} \, , \\[.2cm]
q_{\pa} &\to & \kpa q_{\pa} / p_{\pa} \, . 
\end{array}
\label{trans}
\ee
We search for stationary solutions in the power law form $E(\kpn,\kpa) \sim \kpn^{-n} 
\kpa^{-m}$. (We will only consider positive parallel wavenumber.) The 
new form of the collision integral of equation (\ref{mastersIon2}), resulting 
from the summation of the integrand in its primary form and after the 
Kuznetsov-Zakharov transformation, is 
\be
\partial_t E^V (\kpn,\kpa) = 
\label{mastersIon3}
\ee
$$
-\frac{\epsilon^2}{32} \int 
\sum_{s, s_p, s_q} \sin \psi_{q_{\perp}} (s k_{\perp} + s_p p_{\perp} + s_q q_{\perp})^2 
\left( \frac{s_p p_{\perp} - s_q q_{\perp}}{\kpa \, p_{\perp} \, q_{\perp}} \right)^2
s \kpa s_p p_{\parallel} \, \delta_{\kpa, p_{\pa} q_{\pa}} \, \delta(\Omega_{k,pq})
$$
$$
\kpn^{-n} \kpa^{-m} \, p_{\perp} \, q_{\perp}^{-n} q_{\pa}^{-m} 
\left(1- \left(\frac{p_{\perp}}{\kpn}\right)^{-n-1} 
\left(\frac{p_{\pa}}{\kpa}\right)^{-m}\right)
\left(1-\left(\frac{p_{\perp}}{\kpn}\right)^{2n-5}
\left(\frac{p_{\pa}}{\kpa}\right)^{2m-1}\right)
$$
$$
dp_{\perp} \, dq_{\perp} \, dp_{\pa} \, dq_{\pa} \, . 
$$
The above collision integral vanishes for specific values of $n$ and $m$. The exact 
power law solutions correspond to these values. There are two different kind of 
solutions. The fluxless solution, also called the thermodynamic equilibrium solution, 
correspond to the equipartition state for which the flux of energy is 
zero. For this case, we have
\be
\begin{array}{lll}
n &= &-1 \, , \\[.2cm]
m &= &0 \, . \\[.2cm]
\end{array}
\label{thermo}
\ee
This result can easily be checked by direct substitution in the original wave kinetic
equation. The most interesting solution of the wave kinetic equation (\ref{mastersIon3}) 
is the one for which the flux is non-zero and finite. The exact solution is called 
the Kuznetsov-Zakharov-Kolmogorov (KZK) spectrum and corresponds to the values, 
\be
\begin{array}{lll}
n &= &5/2 \, , \\[.2cm]
m &= &1/2 \, . \\[.2cm]
\end{array}
\label{sol}
\ee
In other words, the KZK solution scales as
\be
E^V(\kpn,\kpa) \sim \kpn^{-5/2} \kpa^{-1/2} \, . 
\label{KZK1}
\ee
We note that the power law scaling found is the same as the one for whistler 
wave turbulence. However here we are dealing with the kinetic energy not the 
magnetic one. A necessary condition for the realizability of the KZK spectra 
is that the turbulence be local in the sense that the behavior of the turbulence 
is determined primarily (but not only) by interaction between wave packets 
of comparable spatial scale (see \eg \cite{balk90}). 
To check {\it a posteriori} the validity of 
the solutions, we need to determine the domain of locality, {\it i.e.}, the 
domain where the collision integral converges. In practice, we check that 
the contribution of nonlocal interactions do not lead to a divergence of 
the collision integral. Here, the condition of locality is automatically 
satisfied since the anisotropic limit introduces in the problem a cut-off which 
prevents any infra-red divergence of the collision integral. This characteristic 
is also observed for internal gravity waves (Caillol and Zeitlin, 2000), inertial 
waves (Galtier, 2003) and whistler waves (Galtier and Bhattacharjee, 2003). 
The other consequence of the existence of such a cut-off is that it is not 
possible to evaluate precisely the 
Kolmogorov constant, \ie the prefactor of the spectra. However, the sign 
of the energy transfer can be computed for a reasonable range of cut-offs. 
We observe a positive sign which means that we have a direct energy cascade. 
Note that this information cannot be obtained through a heuristic reasoning 
but only with a rigorous analysis.

\subsection{Master equations in the limit of pure Alfv\'en wave turbulence}
\label{alfven1}

In the large scale limit ($d_ik \to 0$) for which terms like ${\alfm}^2$ tend to 
$1$, we note that an equipartition between the kinetic and magnetic energies may 
be obtained since their wave kinetic equations tend to be identical: if initially 
there is equipartition ($E^V(\kk)=E^B(\kk)=E(\kk)/2$) then at any time the 
equipartition will be conserved. The large scale limit corresponds to the standard 
MHD approximation for which the equipartition is in fact automatically satisfied 
by the kinematics. The large scale limit of expression (\ref{mastersHall}), for 
a state of equipartition, gives 
\be
\partial_t E (\kk) = 
\label{mastersMHD}
\ee
$$
\frac{\pi \, \epsilon^2}{256} \int 
\sum_{\Lambda, \Lambda_p, \Lambda_q \atop s, s_p, s_q} 
\left( \frac{\sin \psi_k}{k} \right)^2 \, 
(\Lambda k + \Lambda_p p + \Lambda_q q)^2 \, 
\left( s\Lambda k + s_p\Lambda_p p + s_q \Lambda_q q \right)^2
$$
$$
\left( \frac{s_q - s_p}{\kpa} \right)^2 \, 
s \kpa s_p p_{\parallel} \, E(\qq)
\left[ E(\pp) - E(\kk) \right] 
\delta(\Omega_{k,pq})\, \delta_{k,pq} \, d{\bf p} \, d{\bf q} \, ,
$$
with $\Omega_{k,pq}=B_0(s\kpa - s_p p_{\parallel} - s_q q_{\parallel})$. 
As expected, only nonlinear interactions between Alfv\'en waves with different 
directional polarities $s_p=-s_q$ will contribute to the dynamics. After such a 
consideration and an expansion over the polarities, one finds
\be
\partial_t E (\kk) = 
\label{mastersMHD2}
\ee
$$
\frac{\pi \, \epsilon^2}{4 B_0} \int 
\sin^2 \psi_k \, (1+\cos^2\psi_q) \ p^2
E(\qq) \left[ E(\pp) - E(\kk) \right] 
\delta(q_{\parallel})\, \delta_{k,pq} \, d{\bf p} \, d{\bf q} \, ,
$$
where we have used the identities
\be
\sum_{\Lambda, \Lambda_p, \Lambda_q} 
(\Lambda k + \Lambda_p p + \Lambda_q q)^2 
\left( \Lambda k + \Lambda_p p - \Lambda_q q \right)^2 = 
\ee
$$
4 \left[ ((k+p)^2 - q^2)^2 + ((k-p)^2-q^2)^2 \right] =
32 k^2 p^2 (1+\cos^2\psi_q) \, ,
$$
and the well known triangle relations. The angle $\psi_q$ refers to the angle 
opposite to $\qq$ in the triangle defined by $\kk=\pp+\qq$. (The same 
result can actually be found directly from equation (\ref{ke_alfven2}); the 
inversion of the system made in section \ref{mastersequa2} is thus compatible 
with such a limit.) As noted 
before, the large scale limit of Hall MHD leads to the particular case where 
helicities are absent and where shear- and pseudo-Alfv\'en wave energies are the 
same. Equation (\ref{mastersMHD2}) can be recovered from the general kinetic 
equations obtained by Galtier et al. (2000) when the same assumptions are made. 
The proof is given in Appendix \ref{AppenF}. Therefore, all the properties 
previously derived are recovered (scaling laws, direction of the cascade, 
Kolmogorov constant...). In particular, the energy spectrum follows the power
law $E(\kpn,\kpa) \sim \kpn^{-2} f(\kpa)$, where $f$ is an arbitrary function 
that is due to the dynamical decoupling of parallel planes in Fourier space. 
This is an exact solution of the wave kinetic equations.

\section{Anisotropic heuristic description for Hall MHD}
\label{heuris}

The study of wave turbulence in Hall MHD is a difficult task. We have seen above 
that for three specific limits we are able to find the exact power law energy 
spectra. Two of them (Alfv\'en and whistler wave turbulence) were already known 
and a heuristic description was given by Galtier et al. (2000) and Galtier and 
Bhattacharjee (2003). We shall derive here a generalized heuristic description 
able to recover the essential physics underlying the KZK spectra including the 
ion cyclotron wave turbulence. However, note that several phenomenologies 
may be used to take into account the various regimes involving a non trivial 
balance between propagation and nonlinear effects like for standard MHD (see 
\eg Zhou et al., 2004). 

We introduce anisotropy in our description by considering that 
\be
k \approx \kpn \gg \kpa \, .
\ee
The primary variables used in the formalism are the generalized Els\"asser variables 
${\cal Z}_{\Lambda}^s$. Thus the nonlinear time built on the generalized Els\"asser 
variables is 
\be
\tau_{NL} \sim \frac{1}{\kpn \, {\cal Z}_{\Lambda}^s} \, .
\ee
Note here that ${\cal Z}_{\Lambda}^s$ has a dimension of a velocity. In other words, 
it is {\it not} taken in Fourier space as it was introduced in Section \ref{chd}. 
In a similar way, we find the following Hall MHD wave period
\be
\tau_{w} \sim \frac{1}{\ols} = \frac{1}{- B_0 \kpa \, \xi_{\Lambda}^s} \, .
\ee
We introduce now the mean rate of energy dissipation per unit mass
$\Pi$. Contrary to the electron MHD case, we do not have to renormalize 
this quantity since it is automatically taken into account by the generalized 
Els\"asser variables. Then we have
\be
\Pi \sim {E \over \tau_{tr}} \sim {E(\kpn,\kpa) \kpn \kpa \over \tau_{tr}} \, ,
\ee
where the transfer time $\tau_{tr}$ has the usual form given by the 
wave kinetic equation
\be
\tau_{tr} \sim \tau_{NL} {\tau_{NL} \over \tau_w} \, ;
\ee
it gives
\be
\Pi \sim 
{E(\kpn,\kpa) \kpn^3 {{\cal Z}_{\Lambda}^s}^2 \over - B_0 \, \xi_{\Lambda}^s} \, .
\ee
To proceed further, we note that 
\be
({\cal Z}_{\Lambda}^s)^2 \sim (\alf - \alfm)^2 {a_{\Lambda}^s}^2 \, ,
\label{space1}
\ee
and
\be
E \sim (1 + {\alfm}^2) \, {a_{\Lambda}^s}^2 \, .
\label{space2}
\ee
In relations (\ref{space1}) and (\ref{space2}), both ${\cal Z}_{\Lambda}^s$ and 
$a_{\Lambda}^s$ are written in the physical space, {\it not} in Fourier space. 
By using the relationships given in Appendix \ref{relation}, we finally obtain
\be
E(\kpn,\kpa) \sim \sqrt{\Pi B_0} \, \kpn^{-2} \kpa^{-1/2} 
(1 + \kpn^2 d_i^2)^{-1/4} \, .
\label{HeurisE}
\ee
The heuristic prediction proposed here is able to describe anisotropic turbulence 
for the three different limits discussed above. We recover, in the small scale limit 
($\kpn d_i \to \infty$), the expected scaling law for whistler as well as ion cyclotron 
wave turbulence and, in the large scale limit ($\kpn d_i \to 0$), the Alfv\'en wave 
turbulence scaling law (since then the parallel wavenumber is a mute variable). 
The prediction is given for the total energy, therefore for the previous small scale 
limits one needs to consider only the magnetic or the kinetic energy respectively. 
One may understand this point by looking at equation (\ref{zab}) where the relation 
between the generalized Els\"asser variables, the velocity and the magnetic field 
is given. In the small scale limit, we have for whistler waves 
$\xi_{\Lambda}^s \to -s \, d_i k$ and for ion cyclotron waves 
$\xi_{\Lambda}^s \to (-s \, d_i k)^{-1}$ (whereas  in the large scale limit 
$\xi_{\Lambda}^s \to - s$); thus we see that either the kinetic or the magnetic 
field will dominate. Note that we ignore anisotropy and assume $\kpn \sim \kpa \sim k$, 
we arrive at the scaling 
\be
E(k) \sim \sqrt{\Pi B_0} \, k^{-3/2} (1 + k^2 d_i^2)^{-1/4}
\ee 
for the one-dimensional isotropic spectrum from which one recognizes the 
Iroshnikov (1963) and Kraichnan (1965) prediction for MHD. 

\begin{figure}
\centerline{\psfig{figure=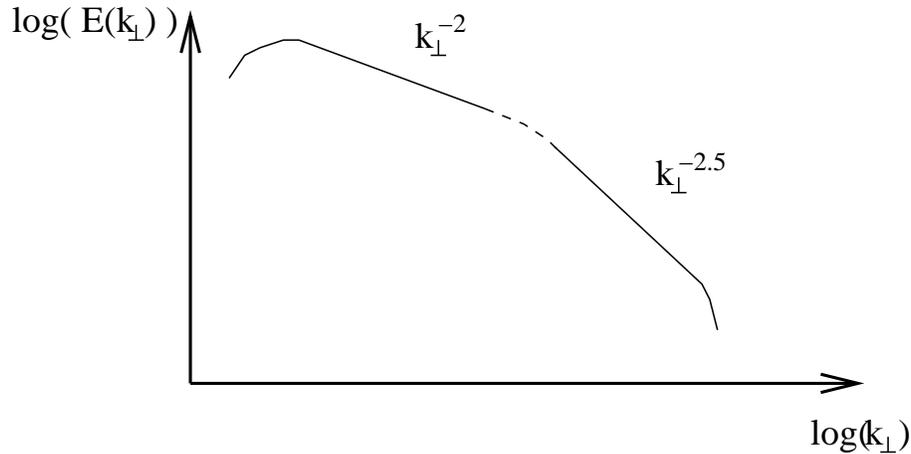,width=12cm,height=6cm}}
\caption{Sketch of the power law spectrum, at a given $\kpa$, expected for 
incompressible Hall MHD in the regime of wave turbulence.}
\label{spectrumeps}
\end{figure}

To summarize our finding we propose the picture given in Fig. \ref{spectrumeps}. 
It is a sketch of the power law energy spectrum predictions for perpendicular 
wavenumbers at a given $\kpa$. The energy spectrum of Hall MHD is characterized 
by two inertial ranges -- the exact power law solutions of the wave kinetic 
equations -- separated by a knee. The position of the knee corresponds to the scale 
where the Hall term becomes sub/dominant, \ie when $\kpn d_i \sim 1$. The heuristic 
prediction tries to make the link continuously (dashed line in Fig. \ref{spectrumeps})
between these power laws but the heuristic spectrum may not be correct at intermediate 
scales since, in particular, Hall MHD turbulence is not necessarily anisotropic. 
However, we will show, in Section \ref{sourceAni}, that even at intermediate scales 
a moderate anisotropy is expected. 
The presence of a knee in the solar wind spectrum is well attested by {\it in situ} 
measurements of magnetic fluctuations (Coroniti et al., 1982; Denskat et al., 1983, 
Leamon et al., 1998; Bale et al., 2005). We will discuss about this point in Section 
\ref{conclusion}; we will also comment the fact the comparison with observations is 
not direct, in particular, for the high frequency part of the spectrum. One reason 
is that the Taylor hypothesis, usually used at low frequency, is not applicable 
anymore. 

\section{Nonlocal interactions}
\label{nonlocality}
\subsection{Introduction}

\begin{figure}[h]
\centerline{\psfig{figure=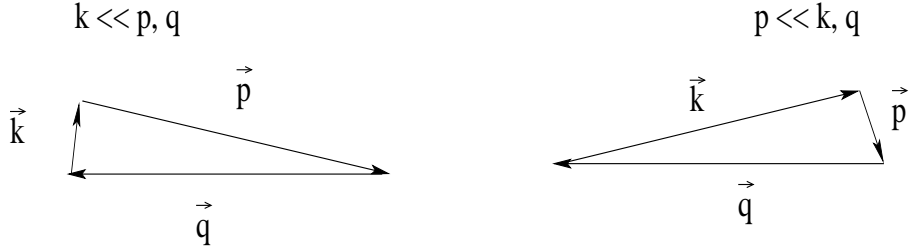,width=12cm,height=3.3cm}}
\caption{Nonlocal triadic interactions.}
\label{nonlocaltriad}
\end{figure}

Nonlocal interactions may play a significant role in turbulence (see \eg Laval 
et al., 2001). By nonlocal interactions, we mean interactions between well-separated 
scales. For triadic processes, it corresponds to the interaction of two highly 
elongated wavevectors with a third small wavevector. In Hall MHD the study of nonlocal 
interactions and nonlocal transfers are particularly interesting since the turbulent 
flow behaves very differently at different scales (see Mininni et al., 2005). At the 
larger scales it follows the MHD dynamics 
whereas at the smallest scales the flow may follow, for example, the EMHD dynamics. 
For simplicity, we will focus our analysis on the master equations (\ref{mastersHall}). 
Therefore we will neglect the helicity effects and keep only the kinetic and magnetic 
energy terms. 
According to the situation, we will consider two different kinds of nonlocal 
interactions, namely $k \ll p, q$ or $p \ll k, q$ (see Fig. \ref{nonlocaltriad}). 
In practice, small scale and large scale effects will be effective when, respectively, 
the limits $k d_i \to +\infty$ and $k d_i \to 0$ will be taken. Therefore, we see that 
large scale effects (small wavenumbers) will be driven by Alfv\'en waves (A) whereas 
small scale effects (large wavenumbers) will be driven by whistler (W) or ion cyclotron 
(C) waves. Note that we will not study nonlocal interactions between the same type of 
waves (AAA, WWW or CCC interactions) which are somewhat more restrictive. Under such a 
condition, three different families of interactions may happen, namely WWA, CCA and 
WCA. We describe below such interactions. 

\subsection{WWA interactions}

First we shall consider the strongly nonlocal interactions of two whistler waves on 
one Alfv\'en wave. In other words, it means that the Alfv\'en wave will be supported 
by the wavevector $\kk$, with $k \ll p, q$. For such a limit the master equations 
(\ref{mastersHall}) reduce to 
\be
\partial_t {E^V (\kk) \brace E^B(\kk)} = 
\label{WWA}
\ee
$$
\frac{\pi \, \epsilon^2}{8 \, d_i^2 B_0^2} \int 
\sum_{\Lambda \atop s, s_p, s_q} 
\left( \frac{\sin \psi_k}{k} \right)^2 \, 
\frac{(\Lambda k + s_p p + s_q q)^2}{2}
\left( \frac{s_q q d_i - s_p p d_i}{\kpa} \right)^2 \, 
{1 \brace 1}
$$
$$
\frac{B_0^2 s\kpa s_p p_{\parallel} p d_i}{2} \, E^B(\qq)
\left[ 
E^B(\pp) - \left( \frac{E^V(\kk) - E^B(\kk)}{-s \Lambda d_i k} \right)
\right] 
\delta(\Omega_{k,pq})\, \delta_{k,pq} \, d{\bf p} \, d{\bf q} \, . 
$$
Note that for whistler waves the polarizations are fixed, \ie $\Lambda_p = s_p$ 
and $\Lambda_q = s_q$. Further simplifications are possible and one obtains finally
\be
\partial_t {E^V (\kk) \brace E^B(\kk)} = 
\label{WWA2}
\ee
$$
\frac{\pi \, \epsilon^2}{8} \int \sum_{s, s_p, s_q} 
\left( \frac{\sin \psi_k}{k} \right)^2 \, 
\left( \frac{s_q q - s_p p}{\kpa} \right)^2 \, 
(s_p p + s_q q) \, \kpa s_p p_{\parallel} p \, {1 \brace 1}
$$
$$
E^B(\qq) \left[ E^V(\kk) - E^B(\kk) \right] 
\delta(\Omega_{k,pq})\, \delta_{k,pq} \, d{\bf p} \, d{\bf q} \, . 
$$
A very interesting result is obtained: at leading order, the Alfv\'en wave is only 
affected by small scale whistler waves when the local (\ie at a given wavevector $\kk$) 
equipartition between the alfv\'enic kinetic and magnetic energies is broken, \ie 
when $E^V(\kk) \neq E^B(\kk)$. 
Additionally, we see that these nonlocal effects are the same for the alfv\'enic 
kinetic and the magnetic energies. This is not so trivial since whistler waves are 
described only by the magnetic energy. We also note that the collisional integral is 
linearly dependent on the difference between the large scale kinetic and magnetic 
energies which can be extracted from the integral. These comments may be summarized 
by the relation
\be
\partial_t {E^V (\kk) \brace E^B(\kk)} = 
I_1(\kk) \, \left[ E^V(\kk) - E^B(\kk) \right] \, {1 \brace 1} \, ,
\label{WWA3}
\ee
where $I_1(\kk)$ is a real function that measures the effects of whistler waves. 

The second case that we shall consider is the one where a whistler wave is affected 
by the interaction between an Alfv\'en wave and another whistler wave. In other words,
it means that the Alfv\'en wave will be supported by the wavevector $\pp$ with 
$p \ll k, q$. The same kind of manipulations as before lead finally to the following 
wave kinetic equations
\be
\partial_t {E^V (\kk) \brace E^B(\kk)} = 
\label{WWA4}
\ee
$$
\frac{\pi \, \epsilon^2}{4} \int \sum_{s, s_p, s_q} 
\left( \frac{\sin \psi_k}{k} \right)^2 \, 
\left( \frac{s k - s_q q}{p_{\parallel}} \right)^2 \, 
(s k + s_q q) \, s \kpa k p_{\parallel} \, {\frac{1}{d_i^2 k^2} \brace 1}
$$
$$
E^B(\qq) \left[ E^B(\pp) - E^V(\pp) \right] 
\delta(\Omega_{k,pq})\, \delta_{k,pq} \, d{\bf p} \, d{\bf q} \, . 
$$
As expected, we note that at leading order the kinetic energy of whistler waves is 
negligible and only the magnetic energy is relevant. We see that the Alfv\'en wave 
affects the whistler wave only when the local equipartition is broken. Thus for any 
kind of WWA nonlocal interactions, one needs a discrepancy from the alfv\'enic 
equipartition to have a non trivial dynamics.

\subsection{CCA interactions}

We study here the strongly nonlocal interactions of two ion cyclotron waves on one 
Alfv\'en wave: in other words, it means that the Alfv\'en wave will be supported by 
the wavevector $\kk$, with $k \ll p, q$. For such a limit the wave kinetic equations 
(\ref{mastersHall}) write
\be
\partial_t {E^V (\kk) \brace E^B(\kk)} = 
\label{CCA}
\ee
$$
\frac{\pi \, \epsilon^2}{8 \, d_i^2 B_0^2} \int 
\sum_{\Lambda \atop s, s_p, s_q} 
\left( \frac{\sin \psi_k}{k} \right)^2 \, 
\frac{(\Lambda k - s_p p - s_q q)^2 \, d_i^8 p^4 q^4}{2 d_i^4 p^2 q^2}
\left( \frac{s_q /q d_i - s_p /p d_i}{\kpa} \right)^2 \, 
{1 \brace 1}
$$
$$
\frac{B_0^2 s\kpa s_p p_{\parallel}}{2 p d_i} \, E^V(\qq)
\left[ 
E^V(\pp) - \left( \frac{E^V(\kk) - E^B(\kk)}{-s \Lambda d_i k} \right)
\right] 
\delta(\Omega_{k,pq})\, \delta_{k,pq} \, d{\bf p} \, d{\bf q} \, . 
$$
For ion cyclotron waves the polarizations are $\Lambda_p = -s_p$ and $\Lambda_q = -s_q$. 
Further simplifications lead to
\be
\partial_t {E^V (\kk) \brace E^B(\kk)} = 
\label{CCA2}
\ee
$$
\frac{\pi \, \epsilon^2}{8 \, d_i^2} \int \sum_{s, s_p, s_q} 
\left( \frac{\sin \psi_k}{k} \right)^2 \, 
\left( \frac{s_q /q - s_p /p}{\kpa} \right)^2 \, (s_p p + s_q q) \, 
p q^2 \kpa s_p p_{\parallel} \, {1 \brace 1}
$$
$$
E^V(\qq) \, \left[ E^B(\kk) - E^V(\kk) \right] 
\delta(\Omega_{k,pq})\, \delta_{k,pq} \, d{\bf p} \, d{\bf q} \, . 
$$
As before, we see that at leading order the Alfv\'en wave is only affected 
by small scale ion cyclotron waves when the local equipartition between the 
alfv\'enic kinetic and magnetic energies is broken. Additionally, we see that these 
nonlocal effects are the same for the alfv\'enic kinetic and magnetic energies. 
This is not so trivial since ion cyclotron waves are described only by the kinetic 
energy. As before, we note that the collisional integral is linearly dependent on 
the difference between the large scale kinetic and magnetic energies which can be 
extracted from the integral. These comments may be summarized by the relation
\be
\partial_t {E^V (\kk) \brace E^B(\kk)} = 
I_2(\kk) \, \left[ E^B(\kk) - E^V(\kk) \right] \, {1 \brace 1} \, ,
\label{CCA3}
\ee
where $I_2(\kk)$ is a real function that measures the effects of ion cyclotron waves. 

The second case that we shall consider is the one where an ion cyclotron wave is 
affected by the interaction between an Alfv\'en wave and another ion cyclotron wave. 
In other words, it means that the Alfv\'en wave will be supported by the wavevector 
$\pp$ with $p \ll k, q$. The same kind of manipulations as before lead finally to 
\be
\partial_t {E^V (\kk) \brace E^B(\kk)} = 
\label{CCA4}
\ee
$$
\frac{\pi \, \epsilon^2}{4 \, d_i^4} \int \sum_{s, s_p, s_q} 
\left( \frac{\sin \psi_k}{k} \right)^2 \, 
\left( \frac{s_q /q - s /k}{p_{\parallel}} \right)^2 \, (s k + s_q q) \, 
\frac{s \kpa p_{\parallel} q^2}{k} \, {d_i^2 k^2 \brace 1}
$$
$$
E^V(\qq) \, \left[ E^V(\pp) - E^B(\pp) \right] 
\delta(\Omega_{k,pq})\, \delta_{k,pq} \, d{\bf p} \, d{\bf q} \, . 
$$
As expected, we note that at leading order the magnetic energy of ion cyclotron waves is 
negligible and only the kinetic energy is relevant. We see that the Alfv\'en wave 
affects the ion cyclotron wave only when its local equipartition is broken. 
Thus for any kind of CCA nonlocal interactions, one needs a discrepancy from the 
alfv\'enic equipartition to obtain a non trivial dynamics.

\subsection{WCA interactions}

First, we investigate the strongly nonlocal interactions of a whistler wave and an ion 
cyclotron wave on an Alfv\'en wave: in this case the Alfv\'en wave will be supported by 
the wavevector $\kk$; the whistler and ion cyclotron waves will be associated respectively 
to $\pp$ and $\qq$, with $k \ll p, q$. For such a limit the wave kinetic equations 
(\ref{mastersHall}) write
\be
\partial_t {E^V (\kk) \brace E^B(\kk)} = 
\label{WCA}
\ee
$$
\frac{\pi \, \epsilon^2}{8 \, d_i^2 B_0^2} \int 
\sum_{\Lambda \atop s, s_p, s_q} 
\left( \frac{\sin \psi_k}{k} \right)^2 \, 
\frac{(\Lambda k + s_p p - s_q q)^2 \, (1 - q^2 / p^2)^2}{2 d_i^2 q^2}
\left( \frac{s_p p d_i - s_q /q d_i}{\kpa} \right)^2 \, 
{1 \brace 1}
$$
$$
\frac{B_0^2 s\kpa s_p p_{\parallel} p d_i}{2} \, E^V(\qq)
\left[ 
E^B(\pp) - \left( \frac{E^V(\kk) - E^B(\kk)}{-s \Lambda d_i k} \right)
\right] 
\delta(\Omega_{k,pq})\, \delta_{k,pq} \, d{\bf p} \, d{\bf q} \, . 
$$
Further simplifications give finally
\be
\partial_t {E^V (\kk) \brace E^B(\kk)} = 
\label{WCA2}
\ee
$$
\frac{\pi \, \epsilon^2}{8 \, d_i^4} \int \sum_{s, s_p, s_q} 
\left( \frac{\sin \psi_k}{k} \right)^2 \, 
\left( \frac{s_p p d_i - s_q /q d_i}{\kpa} \right)^2 \, 
\left( 1 - q^2 / p^2 \right)^2 \, (s_p p - s_q q) \, 
\frac{\kpa s_p p_{\parallel} p}{q^2} \, 
{1 \brace 1}
$$
$$
E^V(\qq) \, \left[ E^V(\kk) - E^B(\kk) \right] 
\delta(\Omega_{k,pq})\, \delta_{k,pq} \, d{\bf p} \, d{\bf q} \, . 
$$
As before, at leading order the Alfv\'en wave is only affected by small scales effects 
when the local alfv\'enic equipartition is broken. These small scale effects are only 
due to ion cyclotron waves. The reason of this apparent asymmetry is actually due 
to the initial association between wavevectors and waves (see below). 
We see that these nonlocal effects are the same for the alfv\'enic kinetic and the 
magnetic energies. Once again, we note that the collisional integral is linearly 
dependent on the difference between the large scale kinetic and magnetic energies 
which can be extracted from the integral. Then one can write
\be
\partial_t {E^V (\kk) \brace E^B(\kk)} = 
I_3(\kk) \, \left[ E^V(\kk) - E^B(\kk) \right] \, {1 \brace 1} \, ,
\label{WCA3}
\ee
where $I_3(\kk)$ is a real function that measures the small scale effects due to 
the ion cyclotron wave. 

Because of the apparent asymmetry mentioned above, we will also consider the 
situation where the whistler wave is associated to the wavevector $\qq$ and the 
ion cyclotron wave to the wavevector $\pp$. For such a limit, we obtain in the same 
way as before the following wave kinetic equations 
\be
\partial_t {E^V (\kk) \brace E^B(\kk)} = 
\label{WCA4}
\ee
$$
\frac{\pi \, \epsilon^2}{8 \, d_i^6} \int \sum_{s, s_p, s_q} 
\left( \frac{\sin \psi_k}{k} \right)^2 \, 
\left( \frac{s_q q d_i - s_p /p d_i}{\kpa} \right)^2 \, 
\left( 1 - p^2 / q^2 \right)^2 \, (s_p p - s_q q) \, 
\frac{\kpa s_p p_{\parallel}}{p^3} \, 
{1 \brace 1}
$$
$$
E^B(\qq) \, \left[ E^B(\kk) - E^V(\kk) \right] 
\delta(\Omega_{k,pq})\, \delta_{k,pq} \, d{\bf p} \, d{\bf q} \, . 
$$
The same comments as before hold: at leading order the Alfv\'en wave is only affected 
by small scales effects when the local alfv\'enic equipartition is broken. Here these 
small scale effects are only due to whistler waves. We also see that these nonlocal 
effects are the same for the alfv\'enic kinetic and the magnetic energies. Once again, 
we note that the collisional integral is linearly dependent on the difference between 
the large scale kinetic and magnetic energies. Then we can write
\be
\partial_t {E^V (\kk) \brace E^B(\kk)} = 
I_4(\kk) \, \left[ E^B(\kk) - E^V(\kk) \right] \, {1 \brace 1} \, ,
\label{WCA5}
\ee
where $I_4(\kk)$ is a real function that measures the small scale effects due to
the whistler wave. 

The next situation that we are going to investigate is the one where an ion cyclotron 
wave is affected by the interaction between an Alfv\'en wave and a whistler wave. 
In other words, it means that the Alfv\'en wave will be supported by the wavevector 
$\pp$, the whistler and ion cyclotron waves will be associated respectively 
to $\qq$ and $\kk$, with $p \ll k, q$. The same type of manipulations as before lead to 
\be
\partial_t {E^V (\kk) \brace E^B(\kk)} = 
\label{WCA5}
\ee
$$
\frac{\pi \, \epsilon^2}{4 \, d_i^8} \int \sum_{s, s_p, s_q} 
\left( \frac{\sin \psi_k}{k} \right)^2 \, 
\left( \frac{s_q q d_i - s /k d_i}{p_{\parallel}} \right)^2 \, 
\left( 1 - k^2 / q^2 \right)^2 \, (s k - s_q q) \, 
\frac{s \kpa p_{\parallel}}{k^5} \, 
{d_i^2 k^2 \brace 1}
$$
$$
E^B(\qq) \, \left[ E^V(\pp) - E^B(\pp) \right] 
\delta(\Omega_{k,pq})\, \delta_{k,pq} \, d{\bf p} \, d{\bf q} \, . 
$$
As expected, at leading order the magnetic energy of ion cyclotron waves is negligible 
and only the kinetic energy is relevant. Once again, the Alfv\'en wave affects the ion 
cyclotron wave only when the local equipartition is broken. 

The last case is the one where a whistler wave is affected by the interaction between 
an Alfv\'en wave and an ion cyclotron wave. In other words, it means that the Alfv\'en 
wave will be supported by the wavevector $\pp$, the whistler and ion cyclotron waves 
will be associated respectively to $\kk$ and $\qq$, with $p \ll k, q$. Under these 
conditions and after some simplifications the master equations write 
\be
\partial_t {E^V (\kk) \brace E^B(\kk)} = 
\label{WCA6}
\ee
$$
\frac{\pi \, \epsilon^2}{4 \, d_i^4} \int \sum_{s, s_p, s_q} 
\left( \frac{\sin \psi_k}{k} \right)^2 \, 
\left( \frac{s_q /q d_i - s k d_i}{p_{\parallel}} \right)^2 \, 
\left( 1 - q^2 / k^2 \right)^2 \, (s k - s_q q) \, 
\frac{s \kpa k p_{\parallel}}{q^2} \, 
{1/d_i^2 k^2 \brace 1}
$$
$$
E^V(\qq) \, \left[ E^B(\pp) - E^V(\pp) \right] 
\delta(\Omega_{k,pq})\, \delta_{k,pq} \, d{\bf p} \, d{\bf q} \, . 
$$
As expected, at leading order the kinetic energy of whistler waves is negligible 
and only the magnetic energy is relevant. Once again, the Alfv\'en wave affects the 
whistler wave only when the local equipartition is broken. In conclusion, for any 
type of WCA nonlocal interactions, one needs a discrepancy from the alfv\'enic 
equipartition to obtain a non trivial dynamics.

\subsection{Discussion}

We have seen that for any kind of nonlocal interactions (WWA, CCA or WCA) a non trivial 
dynamics is present as long as the large scale kinetic and magnetic energies are 
different. This result reveals from the above detail analysis is new and may explain 
some solar wind features. Indeed at first sight, the solar wind may be seen as a MHD 
flow where the kinetic and magnetic energies are balanced. However, fluctuations around 
this equipartition state exist. The present work may explain how these fluctuations are 
maintained by small scales processes. 
Small scales may also be affected by what happens at the largest scales of the system. 
In this case Alfv\'en waves may be seen as a potential source of energy for small scales. 
We report in Appendix \ref{AppenG} the result of the nonlocal analysis made on WCC 
interactions.

\section{Source of anisotropy}
\label{sourceAni}
\subsection{Local interactions}

In previous sections, we have seen three different limits for which anisotropic 
turbulence develops more or less strongly, \ie the small scale limit, with right 
or left polarization, and the large scale limit of Hall MHD. For intermediate 
wavenumbers the situation is more difficult to analyze. Indeed, the wave kinetic 
equations are in general too complex to give a simple answer. In this section, we 
shall see what happens when strongly local interactions are only taken into account. 
It is a situation where wavevectors have approximately the same amplitude, \ie 
$k \approx p \approx q$ (we consider triangles, $\kk = \pp +\qq$, approximately 
equilaterals). We are particularly interested by what happens at intermediate scales. 

Our analyze may start from the master equations (\ref{mastersHall}). The hypothesis 
of strongly local interactions does not lead to drastic simplifications; it seems 
therefore more interesting to look at the resonance condition (\ref{resonance}) 
that appears in equations (\ref{mastersHall}). The first case that we may investigate 
is the one for which waves have the same polarization (left or right). After 
simplifications, one finds the approximate relation
\be
\left(\frac{s_p-s_q}{\kpa}\right)^2 \approx \left(\frac{s_q-s}{p_{\pa}}\right)^2 
\approx \left(\frac{s-s_p}{q_{\pa}}\right)^2 \, . 
\label{reson_local}
\ee
Necessarily two directional wave polarities will be equal which implies that one of 
the parallel wavenumbers will be (approximately) equal to zero. In other words, local 
interactions between the same kind of waves lead to anisotropic turbulence where small 
scales are preferentially generated perpendicular to the external magnetic field. 

The second case that we shall analyze is the one where we have local interactions 
between waves with different polarizations (left and right). Then the resonance 
condition simplifies to
\be
\left(\frac{s_p (s_p \Lambda_p + \sqrt{1+ 4/d_i^2 p^2}) - 
s_q (s_q \Lambda_q + \sqrt{1+ 4/d_i^2 q^2})}{\kpa}\right)^2 \approx 
\label{reson_local2}
\ee
$$
\left(\frac{s_q(s_q \Lambda_q + \sqrt{1+ 4/d_i^2 q^2}) - 
s(s \Lambda + \sqrt{1+ 4/d_i^2 k^2})}{p_{\pa}}\right)^2 \approx 
$$
$$
\left(\frac{s(s \Lambda + \sqrt{1+ 4/d_i^2 k^2}) - 
s_p(s_p \Lambda_p + \sqrt{1+ 4/d_i^2 p^2})}{q_{\pa}}\right)^2 \, . 
$$
Several combinations are possible. We note that for the particular case where 
two waves have both the same polarization (left or right) and the same directional 
polarity, one parallel wavenumber will be (approximately) equal to zero. Thus, 
such a situation leads to the same type of anisotropic turbulence as before. A 
similar analysis applied to all the other cases does not lead to such a conclusion: 
an isotropic turbulence may be generated by the nonlinear dynamics. 

The wave kinetic equations (\ref{mastersHall}) are composed of a sum over the 
different polarities ($\Lambda$- and $s$-type). In other words, all the cases that 
we have mentioned above have to be taken into account to understand what happens 
for strongly local interactions. Both the mechanisms leading to anisotropy and 
isotropy are in competition and therefore it is likely that globally a moderate 
anisotropy (less strong than what happens in the limits of large scale or small 
scale Hall MHD) is generated at intermediate wavenumbers.

\subsection{Discussion}

The local interaction analysis gives some hints to understand globally the 
nonlinear dynamics of wave turbulence in Hall MHD. It is also instructive to 
analyze the opposite limit, namely the limit of strongly nonlocal interactions. 
In Section \ref{nonlocality}, we have introduced the concept of nonlocal 
interactions and discussed about some properties. Information about anisotropy 
can be extracted from the different cases analyzed. More precisely, one needs 
to study the solution of the condition $\Omega_{k,pq}=0$ that has to be satisfied 
at the resonance. The general tendency that we observe (we will not give a detail 
description of every cases) is that most of nonlocal interactions leads to 
anisotropic turbulence where small scales are preferentially generated perpendicular 
to the external magnetic field. 

Both local and nonlocal interactions contribute to develop anisotropy more or less 
strongly. Additionally, we know that a direct energy cascade is predicted in the 
large scale limit (pure Alfv\'en wave turbulence) as well as in the small scale 
limit (whistler and ion cyclotron wave turbulence) of Hall MHD. Thus it is expected 
that a turbulent anisotropic spectrum generated, or developed, at large scale will 
spread out over wavenumbers, from $k d_i < 1$ to $k d_i > 1$, extending anisotropy 
to smaller scales. This scenario may be applicable to the solar wind for which 
various indirect lines of evidence, at low and high frequencies, show that waves 
propagate at large angles to the background magnetic field and that the power in 
fluctuations parallel to the background magnetic field is much less than the 
perpendicular one (Coroniti et al., 1982; Leamon et al., 1998).

\section{Wave versus strong turbulence}
\subsection{Domain of validity of wave turbulence}

Wave turbulence deals with asymptotic developments which are based on a time scale 
separation, with a transfer time assumed to be much larger than the wave period. 
A consequence of this assumption is that the theory is not valid uniformly in all 
of $\kk$-space. In this section, we are going to evaluate the condition of validity 
of the present theory. Generally, one needs to start with the wave kinetic equations 
for which exact power law solutions exist. In our case, different solutions have 
been found for different limits but in general -- for intermediate wavenumbers -- no 
solution has been found. For that reason we are going to use a dimensional 
analysis and, in particular, the heuristic spectrum derived in Section \ref{heuris}. 
This spectrum is based on anisotropic arguments, therefore the condition that we are
going to write is only valid for anisotropic turbulence and might not be applicable 
at intermediate wavenumbers. 

We remind that the nonlinear time and the wave period are respectively
\be
\tau_{NL} \sim \frac{1}{\epsilon \, \kpn {\cal Z}_{\Lambda}^s}
\ee
and
\be
\tau_{w} \sim \frac{1}{\ols} = \frac{1}{- B_0 \kpa \, \xi_{\Lambda}^s} \, .
\ee
Note that we have included explicitly the small parameter $\epsilon$ in the nonlinear 
time. The transfer time $\tau_{tr}$ is built on these time scales and writes
\be
\tau_{tr} \sim \tau_{NL} {\tau_{NL} \over \tau_w} \, . 
\ee
Then the asymptotic condition $\tau_{tr} \gg \tau_{w}$ is 
\be
\frac{-B_0 \kpa \alf}{\epsilon^2 \kpn^2 {\cal Z}_{\Lambda}^s} \gg 
\frac{1}{-B_0 \kpa \alf} \, ,
\label{trcondition}
\ee
where ${\cal Z}_{\Lambda}^s$ has a dimension of a velocity (\ie it is not taken in 
Fourier space as it was introduced in Section \ref{chd}). To proceed further we use 
relations (\ref{space1})--(\ref{space2}) and we take into account the heuristic 
spectrum (\ref{HeurisE}); one obtains (with $B_0 = 1$)
\be
({\cal Z}_{\Lambda}^s)^2 \sim \frac{(\alf - \alfm)^2}{1+{\alfm}^2} 
\kpn^{-1} \kpa^{1/2} (1+\kpn^2 d_i^2)^{-1/4} \, . 
\label{Zcondition}
\ee
Then we insert relation (\ref{Zcondition}) into equation (\ref{trcondition}). 
By using relation (\ref{A13}), we finally obtain the inequality
\be
\kpa \gg \epsilon^{4/3} \kpn^{2/3} (1+ \kpn^2 d_i^2)^{1/2} (1+{\alf}^2)^{-2/3} \, .
\label{thecondition}
\ee
Equation (\ref{thecondition}) is the condition of validity of wave turbulence for 
Hall MHD at the level of three-wave interactions. This condition is {\it not} in 
contradiction with the anisotropic assumption ($\kpn \gg \kpa$) since the small 
parameter $\epsilon$ is present. Thus we can always find a domain of the $\kk$-space 
where the theory is valid. We note, in particular, that for whistler and ion cyclotron
wave turbulence the condition of validity reduces, respectively, to 
$\kpa \gg \epsilon^{4/3} \kpn^{1/3}$ and $\kpa \gg \epsilon^{4/3} \kpn^{5/3}$. 
We remind that it is an evaluation based on anisotropic assumption and therefore it 
might not be valid at intermediate wavenumbers. In the prohibited region (small $\kpa$) 
other higher-order processes such as four-wave interaction have to be taken into 
account. Basically it is the domain where strong turbulence occurs.

\subsection{The two-dimensional state}

The wave turbulence theory developed in this paper describes the Hall MHD dynamics 
of three-dimensional incompressible waves. Indeed, the condition of validity found 
previously shows that the two-dimensional state, \ie modes with $\kpa=0$, is not 
described by the wave kinetic equations. Actually, if we look at equations (\ref{ke1}), 
we see that the nonlinear transfers  decrease (linearly) with $\kpa$. For the 
forbidden value $\kpa=0$ the transfer is exactly null. We see therefore that the 
two-dimensional -- slow -- modes decouple from the three-dimensional Hall MHD waves. 
Such decoupling is not new; it is found in a variety of problems like for internal
gravity waves (see \eg Phillips, 1969), rotating stratified flows (see \eg 
Bartello, 1995) or pure rotating flows (see \eg Smith and Waleffe, 1999). 

As we have already noticed in Section \ref{45} the MHD limit is somewhat singular 
for Hall MHD. A signature of this singular behavior is the appearance of principal 
value terms in the wave kinetic equations. The consequence is that there is an 
important difference between Hall MHD and MHD: in the latter case the two-dimensional 
modes drive the three-dimensional dynamics. Therefore, in this particular case, we 
need information from the forbidden region to describe the dynamics in the domain of 
applicability of wave turbulence. Note that even in this case the wave turbulence 
theory at the level of three-wave interactions has been shown to be applicable 
(Galtier et al., 2000; Nazarenko et al., 2001, Bhattacharjee and Ng, 2001). 
For Hall MHD in general, higher order processes like four-wave interactions may lead 
to a coupling between the two-dimensional state and three-dimensional modes.

\subsection{Strong Hall MHD turbulence}

We shall derive now the equivalent heuristic prediction as in Section \ref{heuris} 
but for strong turbulence. We would like to compare, at least at the level of the 
phenomenology, the wave and strong turbulence regimes. 

We will assume that a moderate external magnetic field $B_0$ is applied and that it 
still affects the Hall MHD dynamics. In other words we will distinguish the wavenumber 
$\kpn$ from $\kpa$, with $\kpn \gg \kpa$, and we suppose that the nonlinear time is 
of the order of the wave period. Then one has
\be
\Pi \sim {E \over \tau_{tr}} \sim {E(\kpn,\kpa) \kpn \kpa \over \tau_{NL}} \, . 
\ee
The equivalent manipulations as in Section \ref{heuris} lead to the prediction
\be
E(\kpn,\kpa) \sim \Pi^{2/3} \, \kpn^{-5/3} \kpa^{-1} 
(1 + \delta_{\Lambda s} \kpn d_i)^{-2/3} \, ,
\label{Heuristrong}
\ee
where $\delta_{\Lambda s}$ is introduced to distinguish the electron MHD case 
($\delta_{ss}=1$) from the ion MHD case ($\delta_{-ss}=0$). The heuristic prediction 
gives the energy spectrum scaling laws for the three different limits of standard, 
electron and ion MHD as well as for intermediate wavenumbers. Contrary to the wave 
turbulence regime the prediction shows that EMHD and IMHD follow different power laws. 
If the external magnetic field is small enough one may ignore anisotropy and write 
$\kpn \sim \kpa \sim k$; it gives 
\be
E(k) \sim \Pi^{2/3} \, k^{-5/3} (1 + \delta_{\Lambda s} k d_i)^{-2/3} \, ,
\label{Heuristrongiso}
\ee
for the one-dimensional isotropic spectrum. It is interesting to note that the 
EMHD prediction is the one found and observed in direct numerical simulations by 
Biskamp et al. (1999). Note that the Kolmogorov scaling found at large scales is an 
unavoidable result since we have used in fact the same phenomenology as Kolmogorov. 
Note finally that the scaling law predictions made for strong and wave turbulence in 
the presence of anisotropy may be described in the context of the generalized critical 
balance proposed by Galtier et al. (2005). In that context the solutions found here 
would be two particular cases of a familly of solutions. 
The result obtained here has to be contrasted with those obtained by Krishan and 
Mahajan (2004,2005) for strong isotropic turbulence. In these papers, the authors give 
several predictions separately for the magnetic and kinetic energy spectra. 
In particular, they conclude that the whistler-type prediction is very different 
from the observations and evoke a possible stronger damping for whistler waves. In the 
present paper, it is thought that the damping (\eg due to the resonance) affects mainly 
the ion cyclotron waves (see also Goldstein et al., 1994).

\section{Conclusion}
\label{conclusion}

\subsection{Summary}

In this paper we have investigated the steepening of the magnetic fluctuation power 
law spectra observed in the solar wind for frequencies higher than $0.5$~Hz. We have 
shown that the high frequency part of the spectrum may be attributed to dispersive 
nonlinear processes present in the incompressible Hall MHD equations rather than 
pure dissipation. In that context, we have introduced the generalized Els\"asser 
variables adapted to Hall MHD and developed a wave turbulence formalism based on a 
complex helical 
decomposition. Then, we have derived the wave kinetic equations for three dimensional 
incompressible Hall MHD turbulence at the level of three-wave interactions, in the 
presence of a strong external magnetic field, and in the most general case, \ie 
with global and magnetic helicities. In such a regime, Hall MHD turbulence is 
mainly anisotropic with small scales preferentially generated perpendicular to the 
external magnetic field. In particular, strong anisotropy is developed at small 
scales (ion cyclotron/IMHD and whistler/EMHD wave turbulence) and large scales (pure 
Alfv\'en/MHD wave turbulence). For intermediate wavenumbers, \ie  $k d_i \approx 1$ 
only a moderate anisotropy may happen. We have found two different power law energy 
spectra that are exact solutions of the wave kinetic equations, with a steeper power 
law at smaller scales. To illustrate this finding, we have developed an anisotropic 
phenomenology that describes continuously the different scaling laws for the energy 
spectrum: the existence of a double inertial range is recovered with a knee at 
intermediate scales where the Hall term becomes (sub-)dominant. We have also shown 
that the large scale limit of Hall MHD, \ie the standard MHD, is singular 
and leads to the appearance of principal value terms in the wave kinetic equations. 
We have analyzed the nonlocal interactions between Alfv\'en, whistler and ion 
cyclotron waves and we have shown that, at leading order, a non trivial dynamics 
is present only when a discrepancy from the equipartition between the large scale 
kinetic and magnetic energies of Alfv\'en waves happens. Finally, we have given 
the conditions of validity of wave turbulence in Hall MHD and compared the wave 
and strong turbulence scaling laws.

\subsection{Discussion}

Solar wind magnetic fluctuation power law spectra are commonly observed to have 
an index around $-5/3$ at frequencies lower than $0.5$Hz and a steeper power 
law behavior at higher frequencies with a spectral index on average around $-3$ 
(Coroniti et al., 1982; Denskat et al., 1983, Leamon et al., 1998). The latest 
analysis made from Cluster spacecraft data has -- however -- reported a spectral 
index about $-2.12$ (Bale et al., 2005). These small scale fluctuations are also 
characterized by a bias of the polarization suggesting that they are likely to 
be right-hand polarized, outward propagating waves (Goldstein et al., 1994). For 
these reasons, it is claimed that Alfv\'en -- left circularly polarized -- 
fluctuations may be suppressed by proton cyclotron 
damping and that the high frequency power law spectra are likely to consist of 
whistler waves (see \eg Stawicki et al., 2001). Additionally, it is often observed 
that the high frequency fluctuations of the magnetic field are much smaller than 
the background magnetic field (see \eg Stawicki et al., 2001). Therefore, a wave 
turbulence approach seems to be well adapted to this problem. It is well known that 
whistler waves are solution to the Hall MHD equations. We have seen in the regime 
of wave turbulence that when the Hall term is added to the MHD equations a 
nontrivial dynamics happens at small scales (\ie when $k d_i > 1$) with, in 
particular, a steepening of the power law energy spectra. This behavior is similar 
to what is observed in the solar wind and may be seen as an indication that the 
steepening of the solar wind magnetic fluctuation power law spectra is mainly due 
to nonlinear processes rather than pure dissipation. Under this new interpretation, 
the resistive dissipation range of frequencies may be moved to frequencies higher 
than the electron cyclotron frequency. 

The Taylor hypothesis, that allows to connect directly a frequency to a wavenumber, 
is widely used to interpret the single spacecraft solar wind data. It is thought 
that this approximation is well adapted to analyze, in particular, the low frequency 
part of the magnetic and velocity fluctuations. The subsequent interpretation is 
mainly relevant for isotropic media. However, there are evidences that anisotropy 
is present in the solar wind at high and low frequencies (see \eg Belcher and Davis, 
1971). From a theoretical point of view, it is well known that the presence of a 
strong magnetic field influences the MHD turbulent flows (see \eg Pouquet, 1978; 
Montgomery and Turner, 1981; Shebalin et al. 1983; Goldreich and Sridhar, 1995; Ng 
and Bhattacharjee, 1996-1997; Galtier et al., 2000-2005). The main effect is that 
MHD turbulence becomes mainly bidimensional with a nonlinear transfer essentially 
perpendicular to its direction. Direct numerical simulations of $2{1 \over 2}$D 
compressible Hall MHD for high and low beta $\beta$ plasma (Ghosh and Goldstein, 
1997) have also displayed such an anisotropic property when a strong magnetic field 
is present. As we have seen, the model that we propose here is also able to exhibit 
such an anisotropy. The predictions that we have made are for anisotropic turbulence, 
\ie a situation where we distinguish the wavenumber $\kpn$ from $\kpa$. Therefore, 
a proper comparison with observational data will be possible only when a three 
dimensional energy spectrum will be accessible. It is important to note that average 
effects may alter significantly the power law scaling. An illustration of such 
effects may be given from the recent heuristic MHD predictions made by \cite{Galtier05} 
who have generalized the concept of critical balance. For a medium where a strong or 
moderate magnetic field $B_0 \ep$ is present, they predict the anisotropic energy 
spectrum $E(\kpn,\kpa) \sim \kpn^{-\alpha} \kpa^{-\beta}$, with $3\alpha+2\beta=7$. 
This model is able to describe both the strong and wave turbulence regimes as well 
as the transition between them; it also satisfies the critical balance relationship 
$\kpa \propto \kpn^{2/3}$. It is very interesting to note that if one averages the 
heuristic spectrum to obtain the one dimensional counterpart $E(\kpn)$, one obtains 
systematically a scaling law in $\kpn^{-5/3}$ for any family of solutions ($\alpha$,
$\beta$). Actually, this remark might explain the apparent contradiction between, 
on the one hand, the presence of anisotropy in the solar wind and, on the other hand,
the Kolmogorov ($-5/3$) scaling law for the low frequency energy spectrum. 
Of course, other explanations of the $5/3$--spectrum are possible; an example is 
given in Oughton and Matthaeus (2005). 
The direct comparison between theoretical predictions and {\it in situ} measurements 
of three dimensional energy spectra is particularly crucial for the high frequency 
part of the power spectrum for which the usual Taylor hypothesis that allows to 
connect directly the frequency to the wavenumber is not applicable anymore. 
Efforts are currently made with Cluster spacecraft data from which it is possible 
to extract the three dimensional magnetic turbulent spectra of the magnetosheath 
thanks to multipoint measurements and a k-filtering technique (see \eg Sahraoui 
et al., 2004). As explained in \cite{Bale}, Cluster spacecraft may exit from the 
terrestrial magnetosphere to make solar wind measurements. The application of the 
k-filtering technique to the high frequency part of the solar wind magnetic 
fluctuations seems then possible. It may lead, for the first time, to a direct 
and rigorous comparison with a model prediction of solar wind. Note that using 
multispacecraft analysis, Matthaeus et al. (2005) have recently investigated some 
spatial correlations from two-point measurements. 

The model proposed in this paper is able to recover some solar wind properties but 
several aspects have not been discussed. For example, it would be interesting to 
investigate the effects of asymmetry, \ie the fact that outward propagating waves 
dominate in the solar wind. In the framework of incompressible MHD, we know that 
asymmetry changes the index of the power law spectra. Similarly, the indices found 
here may be affected by asymmetry; this question will be tackled in the future. 
In this paper, we have found that a turbulent state made of ion cyclotron waves 
may exist around a fraction of the ion cyclotron frequency $\omega_{ci}$, namely 
for a frequency around $\omega_R = (\kpa / k) \, \omega_{ci}$. The presence of a 
resonance at a frequency $\omega_R$ lower than $\omega_{ci}$ is rarely discussed 
in the litterature (see White et al., 2002) since the analysis 
is generally focused on parallel propagations ($\kpa = k$) for which the resonance 
frequency is exactly the ion cyclotron frequency. As we have seen, in the presence 
of a strong external magnetic field, Hall MHD turbulence becomes mainly bidimensional 
with an energy spectrum mainly spreaded out over the wavenumbers $\kpn$. In this case, 
the resonance frequency may appear at a small fraction of $\omega_{ci}$ (this fraction 
being even smaller for heavier mass ions). These remarks 
may be crucial to understand the solar coronal heating problem in which the coronal 
temperature is far beyond what one can predict by the resistive MHD approximation:
although the Hall MHD model is a fluid model that does not describe the resonance 
between waves and particules and therefore the particule heating, it offers the 
possibility to evaluate the rate of particule heating by assuming that the turbulent 
energy of ion cyclotron waves are mainly transfer into heating. This point is currently 
investigated and will be reported elsewhere. In the context of the solar wind, the 
measure of the position of the knee in the magnetic fluctuation power law spectrum 
may be seen as a proxy to measure the solar wind anisotropy. 

The Hall effect is relevant in many astrophysical problems to understand, for example, 
the presence of instabilities in protostellar disks (Balbus and Terquem, 2001), 
the magnetic field evolution in neutron star crusts (Goldreich and Reisenegger, 
1992; Cumming et al, 2004) or impulsive magnetic reconnection (see \eg 
Bhattacharjee, 2004). Small scale turbulence is also a key issue in a number 
of problems from the interstellar medium (see the review made by \cite{elmegreen},
and, \cite{scalo}), to solar physics (see \eg, \cite{petrosyan}), magnetospheric 
physics (see \eg the recent paper by \cite{Goldstein05}) and laboratory devices 
such as tokamaks (see \eg Wild et al., 1981; Taylor, 1993). Therefore, it is likely 
that the present model will interest several other (astrophysical) problems. 

\bigskip

\section*{acknowledgments}
I would like to thank G\'erard Belmont, Peter Goldreich, Pablo Mininni and Fouad 
Sahraoui for useful discussions. The author acknowledge partial financial support 
from the PNST (Programme National Soleil--Terre) program of INSU (CNRS) and from 
the Research Training Network ``Theory, Observations and Simulations of Turbulence 
in Space Plasmas'' through European Community grant HPRN-CT-2001-00310. 

\newpage

\appendix
\section{Some useful relationships}
\label{relation}

It is convenient to note the following identities: 
\be
1-{\alf}^2 = k \Lambda \, d_i \, \alf \, ,
\ee
\be
\alf \, \alfm = -1 \, ,
\ee
\be
\alf - \alfm = - s \sqrt{4+k^2 d_i^2} \, ,
\label{A13}
\ee
\be
\alf + \alfm = - k \, d_i \, \Lambda \, ,
\ee
\be
\Lambda \left(\frac{1-{\alfm}^4}{k}\right) \left(\frac{\ols}{1+{\alfm}^2}\right) = 
d_i \, \kpa \, .
\label{ident5}
\ee

\section{Derivation of the wave kinetic equations}
\label{derivation}

The starting point of the derivation of the wave kinetic equations for incompressible
Hall MHD is the fundamental equation (\ref{fonda2}). We write successively equations 
for the second and third-order moments,
\be
\partial_t \langle \aak \aakp \rangle = 
\label{second}
\ee
$$
\frac{\epsilon}{4 \, d_i}
\int \sum_{\Lambda_p, \Lambda_q \atop s_p, s_q}
{\alf}^2 \, \frac{\alfq - \alfp}{\alf - \alfm} \, 
M{{\Lambda \Lambda_p \Lambda_q \atop s \, s_p \, s_q} \atop -k \, p \, q}
\, \langle \aap \aaq \aakp \rangle \, e^{-i \Omega_{pq,k} t} \, \delta_{pq,k} \, 
d{\bf p} \, d{\bf q}
$$
$$
+
$$
$$
\frac{\epsilon}{4 \, d_i}
\int \sum_{\Lambda_p, \Lambda_q \atop s_p, s_q}
{\alfprime}^2 \, \frac{\alfq - \alfp}{\alfprime - \alfmprime} \, 
M{{\Lambda^{\prime} \Lambda_p \Lambda_q \atop s^{\prime} \, s_p \, s_q} 
\atop -k^{\prime} \, p \, q}
\, \langle \aap \aaq \aak \rangle \, e^{-i \Omega_{pq,k^{\prime}} t} \, 
\delta_{pq,k^{\prime}} \, d{\bf p} \, d{\bf q} \, ,
$$
and 
\be
\partial_t \langle \aak \aakp \aakdp \rangle = 
\label{third}
\ee
$$
\frac{\epsilon}{4 \, d_i}
\int \sum_{\Lambda_p, \Lambda_q \atop s_p, s_q}
{\alf}^2 \, \frac{\alfq - \alfp}{\alf - \alfm} \, 
M{{\Lambda \Lambda_p \Lambda_q \atop s \, s_p \, s_q} \atop -k \, p \, q}
\, \langle \aap \aaq \aakp \aakdp \rangle \, e^{-i \Omega_{pq,k} t} \, \delta_{pq,k} \, 
d{\bf p} \, d{\bf q}
$$
$$
+
$$
$$
\frac{\epsilon}{4 \, d_i}
\int \sum_{\Lambda_p, \Lambda_q \atop s_p, s_q}
{\alfprime}^2 \, \frac{\alfq - \alfp}{\alfprime - \alfmprime} \, 
M{{\Lambda^{\prime} \Lambda_p \Lambda_q \atop s^{\prime} \, s_p \, s_q} \atop 
-k^{\prime} \, p \, q}
\, \langle \aap \aaq \aak \aakdp \rangle \, e^{-i \Omega_{pq,k^{\prime}} t} \, 
\delta_{pq,k^{\prime}} \, d{\bf p} \, d{\bf q}
$$
$$
+
$$
$$
\frac{\epsilon}{4 \, d_i}
\int \sum_{\Lambda_p, \Lambda_q \atop s_p, s_q}
{\alfdprime}^2 \, \frac{\alfq - \alfp}{\alfdprime - \alfmdprime} \, 
M{{\Lambda^{\prime \prime} \Lambda_p \Lambda_q \atop s^{\prime \prime} \, s_p \, s_q} 
\atop -k^{\prime \prime} \, p \, q} \, \langle \aap \aaq \aak \aakp \rangle \, 
e^{-i \Omega_{pq,k^{\prime \prime}} t} \, \delta_{pq,k^{\prime \prime}} \, d{\bf p} \, 
d{\bf q}\, .
$$
We shall write asymptotic closure (Newell et al.,  2001) for our system. For that, 
we basically need to write the fourth-order moment in terms of a sum of the fourth-order 
cumulant plus products of second order ones. The asymptotic closure depends on two 
ingredients: the first is the degree to which the linear waves interact to randomize 
phases; the second relies on the fact that the nonlinear regeneration of the third-order 
moment by the fourth-order moment in equation (\ref{third}) depends more on the product of 
the second order moments than it does on the fourth order cumulant. The fourth--order 
moment decomposes into the sum of three products of second--order moments, and a 
fourth--order cumulant. The latter does not contribute to secular behavior, and among 
the other products one is absent because of the homogeneity assumption. If we use the 
symmetric relations (\ref{prop1})--(\ref{prop4}) and perform wavevector integrations, 
summations over polarities and time integration, then equation (\ref{third}) becomes:
\be
\langle \aak \aakp \aakdp \rangle = 
\frac{\epsilon}{4 \, d_i} \, \Delta(\Omega_{k k^{\prime} k^{\prime \prime}}) \, 
\delta_{k k^{\prime} k^{\prime \prime}}
\label{third2}
\ee
$$
\{ \, 
{\alf}^2 \left[\frac{\alfdprime - \alfprime}{\alf - \alfm} 
\left(M{{\Lambda \Lambda^{\prime} \Lambda^{\prime \prime} \atop s \, s^{\prime} \, 
s^{\prime \prime}} \atop k \, k^{\prime} \, k^{\prime \prime}}\right)^* + 
\frac{\alfprime - \alfdprime}{\alf - \alfm} 
\left(M{{\Lambda \Lambda^{\prime \prime} \Lambda^{\prime} \atop 
s \, s^{\prime \prime} \, s^{\prime}} \atop k \, k^{\prime \prime} \, k^{\prime}}\right)^*
\right] \qlsprime \, \qlsdprime
$$
$$
+
$$
$$
{\alfprime}^2 \left[\frac{\alfdprime - \alf}
{\alfprime - \alfmprime} 
\left(M{{\Lambda^{\prime} \Lambda \Lambda^{\prime \prime} \atop s^{\prime} \, s \, 
s^{\prime \prime}} \atop k^{\prime} \, k \, k^{\prime \prime}}\right)^* + 
\frac{\alf - \alfdprime}{\alfprime - \alfmprime} 
\left(M{{\Lambda^{\prime} \Lambda^{\prime \prime} \Lambda \atop 
s^{\prime} \, s^{\prime \prime} \, s} \atop k^{\prime} \, k^{\prime \prime} \, k}\right)^*
\right] \qls \, \qlsdprime
$$
$$
+
$$
$$
{\alfdprime}^2 
\left[\frac{\alf - \alfprime}{\alfdprime - \alfmdprime} 
\left(M{{\Lambda^{\prime \prime} \Lambda^{\prime} \Lambda 
\atop s^{\prime \prime} \, s^{\prime} \, s} 
\atop k^{\prime \prime} \, k^{\prime} \, k}\right)^* + 
\frac{\alfprime - \alf}{\alfdprime - \alfmdprime} 
\left(M{{\Lambda^{\prime \prime} \Lambda \Lambda^{\prime} \atop 
s^{\prime \prime} \, s \, s^{\prime}} \atop k^{\prime \prime} \, k \, k^{\prime}}\right)^*
\right] \qlsprime \, \qls \, \} \, ,
$$
where 
\be
\Delta(\Omega_{k k^{\prime} k^{\prime \prime}})= 
\int_0^t e^{i \Omega_{k k^{\prime} k^{\prime \prime}} t^{\prime}} dt^{\prime} 
= {e^{i \Omega_{k k^{\prime} k^{\prime \prime}}t} - 1 
\over i \Omega_{k k^{\prime} k^{\prime \prime}}} \, .
\ee
The introduction of symmetric relations (\ref{prop1})--(\ref{prop4}) into (\ref{third2}) 
allows us to simplify further the previous equation; one obtains:
\be
\langle \aak \aakp \aakdp \rangle = 
\frac{\epsilon}{2 \, d_i} \, \Delta(\Omega_{k k^{\prime} k^{\prime \prime}}) \, 
\delta_{k k^{\prime} k^{\prime \prime}}
\left(M{{\Lambda \Lambda^{\prime} \Lambda^{\prime \prime} \atop s \, s^{\prime} \, 
s^{\prime \prime}} \atop k \, k^{\prime} \, k^{\prime \prime}}\right)^*
\label{third3}
\ee
$$
\left[
{\alf}^2 \frac{\alfdprime - \alfprime}{\alf - \alfm} \qlsprime \, \qlsdprime + 
{\alfprime}^2 \frac{\alf - \alfdprime}{\alfprime - \alfmprime} \qls \, \qlsdprime + 
{\alfdprime}^2 \frac{\alfprime - \alf}{\alfdprime - \alfmdprime} \qls \, \qlsprime
\right] \, .
$$
We insert expression (\ref{third3}) into equation (\ref{second}); it leads to: 
\be
\partial_t \qls (\kk) = 
\label{secondbis}
\ee
$$
\frac{\epsilon^2}{8 \, d_i^2}
\int \sum_{\Lambda_p, \Lambda_q \atop s_p, s_q}
{\alf}^2 \, \frac{\alfq - \alfp}{\alf - \alfm} \, 
\left|M{{\Lambda \Lambda_p \Lambda_q \atop s \, s_p \, s_q} \atop -k \, p \, q}\right|^2
\Delta(\Omega_{p q, k}) \, e^{- i \Omega_{p q, k} t} \, \delta_{pq,k}
$$
$$
\left[
{\alfp}^2 \frac{\alf - \alfq}{\alfp - \alfmp} \qls \, \qlsq + 
{\alfq}^2 \frac{\alfp - \alf}{\alfq - \alfmq} \qls \, \qlsp + 
{\alf}^2 \frac{\alfq - \alfp}{\alf - \alfm} \qlsp \, \qlsq
\right] 
d{\bf p} \, d{\bf q}
$$
$$
+
$$
$$
\frac{\epsilon^2}{8 \, d_i^2}
\int \sum_{\Lambda_p, \Lambda_q \atop s_p, s_q}
{\alfprime}^2 \, \frac{\alfq - \alfp}{\alfprime - \alfmprime} \, 
\left|M{{\Lambda^{\prime} \Lambda_p \Lambda_q \atop s^{\prime} \, s_p \, s_q} 
\atop -k^{\prime} \, p \, q}\right|^2
\Delta(\Omega_{p q, k^{\prime}}) \, e^{- i \Omega_{p q, k^{\prime}} t} \, 
\delta_{pq,k^{\prime}}
$$
$$
\left[
{\alfp}^2 \frac{\alfprime - \alfq}{\alfp - \alfmp} \qlsprime \, \qlsq + 
{\alfq}^2 \frac{\alfp - \alfprime}{\alfq - \alfmq} \qlsprime \, \qlsp + 
{\alfprime}^2 \frac{\alfq - \alfp}{\alfprime - \alfmprime} \qlsp \, \qlsq
\right] 
d{\bf p} \, d{\bf q} \, .
$$
The long-time behavior of the wave kinetic equation (\ref{secondbis}) is given by 
the Riemman-Lebesgue Lemma which tells us that, for $t \to +\infty$, we have
\be
e^{-ix t} \Delta(x) = \Delta(-x) \to \pi \delta(x) - i {\cal P}(1/x) \, ,
\ee
where ${\cal P}$ is the principal value of the integral. The two terms of equation 
(\ref{secondbis}) are complex conjugated therefore if in the second term we replace 
the dummy integration variables $\pp$, $\qq$, by $-\pp$, $-\qq$, we can simplify 
further equation (\ref{secondbis}) since, in particular, principal value terms 
compensate exactly. Finally, we obtain the wave kinetic equations for incompressible 
Hall MHD:
\be
\partial_t \qls (\kk) = 
\label{ke0}
\ee
$$
\frac{\pi \, \epsilon^2}{4 \, d_i^2}
\int \sum_{\Lambda_p, \Lambda_q \atop s_p, s_q}
\alf \, \frac{\alfq - \alfp}{1 - {\alfm}^2} \, 
\left|M{{\Lambda \Lambda_p \Lambda_q \atop s \, s_p \, s_q} \atop -k \, p \, q}\right|^2
\delta(\Omega_{k, p q}) \, \delta_{k,pq}
$$
$$
\left[
\alfp \frac{\alf - \alfq}{1 - {\alfmp}^2} \, \qls \, \qlsq + 
\alfq \frac{\alfp - \alf}{1 - {\alfmq}^2} \, \qls \, \qlsp + 
\alf \frac{\alfq - \alfp}{1 - {\alfm}^2} \, \qlsp \, \qlsq
\right] 
d{\bf p} \, d{\bf q} \, , 
$$
where
$$
\left|M{{\Lambda \Lambda_p \Lambda_q \atop s \, s_p \, s_q} \atop -k \, p \, q}\right|^2 =
\left( \frac{\sin \psi_k}{k} \right)^2 \, (\Lambda k + \Lambda_p p + \Lambda_q q)^2 \, 
\left( 1 - {\alfm}^2 {\alfmp}^2 {\alfmq}^2 \right)^2 \, 
\, .
$$
The last step that we have to follow to obtain the same expression as (\ref{ke1}) is to 
include the resonance relations (\ref{resonance}) into the previous equations.

\section{Pseudo-dispersive MHD waves and complex helical decomposition}
\label{AppenD}

This section is devoted to the derivation of the wave kinetic equations for
pure alfv\'enic turbulence ($kd_i=0$). These equations were already derived 
by Galtier et al. (2000) but here we use the complex helicity decomposition.
We will see that some differences appear in the kinematics that renders the 
MHD description somewhat singular (principal value terms appear) by opposition 
to the dispersive Hall MHD description. 
We start from the standard incompressible and inviscid MHD equations, 
\be
\nabla \cdot {\bf V} = 0 \, ,
\label{mhd1b}
\ee
\be
\frac{\partial {\bf V}}{\partial t} + {\bf V} \cdot \nabla \, {\bf V} = 
- {\bf \nabla} P_* + \bB \cdot \nabla \, \bB \, ,
\label{mhd2b}
\ee
\be
\frac{\partial \bB}{\partial t} + {\bf V} \cdot \nabla \, \bB = 
\bB \cdot \nabla \, {\bf V} \, ,
\label{mhd3b}
\ee
\be
\nabla \cdot \bB = 0 \, . 
\label{mhd4b}
\ee
The same notation as before is used. We introduce the fluctuating fields
$\bB ({\bf x}) = B_0 \, \ep + \epsilon \, {\bf b} ({\bf x})$, 
${\bf V} ({\bf x}) = \epsilon \, {\bf v} ({\bf x})$ and we Fourier 
transform the MHD equations. One obtains:
\be
\partial_t {\bf z^s}_\kk - i s \kpa B_0 {\bf z^s}_\kk = 
- \epsilon \, \{ {\bf z^{-s}} \cdot \nabla \, {\bf z^s} - {\bf \nabla} P_* \}_\kk \, ,
\label{wave11}
\ee
\be
\kk \cdot {\bf z^s}_\kk = 0 \, ,
\label{wave12}
\ee
where we use the standard Els\"asser variables ${\bf z^s} = {\bf v} + s {\bf b}$. 
The linear solution ($\epsilon=0$) corresponds to linearly polarized Alfv\'en waves 
for which $\omega_k = B_0 \kpa$. We introduce the complex helicity basis and define
\be
{\bf z}^s_\kk = \sum_{\Lambda} \, {\cal Z}^s_{\Lambda} (\kk) e^{i s \omega_k t} \, 
{\bf h^{\Lambda}_k} = 
\sum_{\Lambda} \, {\cal Z}^s_{\Lambda} e^{i s \omega_k t} \, {\bf h^{\Lambda}_k} \, . 
\label{zz}
\ee
We see that this decomposition is not natural for linear polarized Alfv\'en waves 
since for a given direction of propagation (a given $s$) we have two contributions 
for each value of $\Lambda$. We nevertheless use this decomposition to show the 
compatibility with the Hall MHD turbulence description. 
After the substitution of (\ref{zz}) into (\ref{wave11})--(\ref{wave12}), we obtain
\be
\partial_t {\cal Z}^s_{\Lambda} = 
- \frac{i \epsilon}{2} \int \sum_{\Lambda_p, \Lambda_q} 
{\cal Z}^{-s}_{\Lambda_p} \, {\cal Z}^s_{\Lambda_q} 
(\kk \cdot {\bf h^{\Lambda_p}_p}) ({\bf h^{\Lambda_q}_q} \cdot {\bf h^{-\Lambda}_k})
\, 
e^{-i s (\omega_k+\omega_p-\omega_q) t} \, \delta_{pq,k} \, d{\bf p} \, d{\bf q} \, .  
\ee
Note that we do not have a summation over the directional polarity $s_p$ and $s_q$. 
The physical reason is that the nonlinear coupling in incompressible MHD involves 
only Alfv\'en waves propagating in opposite directions. Thus the 
information about the direction of propagation is already taken into account in 
equation (\ref{wave11}). We use the local decomposition and find after some algebra
\be
\partial_t {\cal Z}^s_{\Lambda} = 
\frac{\epsilon}{4} \int \sum_{\Lambda_p, \Lambda_q}
(k\Lambda + q\Lambda_q - p\Lambda_p) \, 
N^{\Lambda \Lambda_p \Lambda_q}_{-k \, p \, q} \, 
{\cal Z}^{-s}_{\Lambda_p} \, {\cal Z}^s_{\Lambda_q} \, 
e^{-2 i s B_0 p_{\parallel} t} \, \delta_{pq,k} \, d{\bf p} \, d{\bf q} \, ,
\label{fonda21}
\ee
where
\be
N^{\Lambda \Lambda_p \Lambda_q}_{k \, \, p \, \, q} = 
e^{i (\Lambda \Phi_k + \Lambda_p \Phi_p + \Lambda_q \Phi_q)} \, 
\Lambda \, \Lambda_p \, \Lambda_q \, \frac{\sin \psi_k}{k} \, 
(\Lambda k + \Lambda_p p + \Lambda_q q) \, .
\label{fonda21bis}
\ee
We note that the matrix $N$ possesses the following properties ($*$ denotes the 
complex conjugate),
\be
\left(
N^{\Lambda \Lambda_p \Lambda_q}_{k \, \, p \, \, q} \right)^* = 
N^{-\Lambda -\Lambda_p -\Lambda_q}_{\, \, \, \, \, k \, \, \, \, \, \, 
p \, \, \, \, \, \, \, \, q} =
N^{\Lambda \, \, \, \, \Lambda_p \, \, \, \, \Lambda_q}_{-k-p-q} \, , 
\label{propz1}
\ee
\be
N^{\Lambda \Lambda_p \Lambda_q}_{k \, \, p \, \, q} = 
- N^{\Lambda \Lambda_q \Lambda_p}_{k \, \, q \, \, p} \, ,
\label{propz2}
\ee
\be
N^{\Lambda \Lambda_p \Lambda_q}_{k \, \, p \, \, q} = 
- N^{\Lambda_q \Lambda_p \Lambda}_{q \, \, p \, \, k} \, ,
\label{propz3}
\ee
\be
N^{\Lambda \Lambda_p \Lambda_q}_{k \, \, p \, \, q} = 
-N^{\Lambda_p \Lambda \Lambda_q}_{p \, \, k \, \, q} \, .
\label{propz4}
\ee
Equation (\ref{fonda21}) is the fundamental equation that describes the slow evolution 
of the Alfv\'en wave amplitudes due to the nonlinear terms of the incompressible MHD 
equations. Note that (i) we have already used the resonance condition to simplify the 
coefficient in the complex exponential function; (ii) a comparison with the large scale 
limit of equation (\ref{fonda2}) is possible if we sum over the directional polarities. 
We follow the same steps as in Appendix \ref{derivation}: we write successively 
equations for the second and third-order moments,
\be
\partial_t 
\langle 
{\cal Z}^s_{\Lambda} {\cal Z}^{s^{\prime}}_{\Lambda^{\prime}} 
\rangle = 
\label{secondmhd}
\ee
$$
\frac{\epsilon}{4}
\int \sum_{\Lambda_p, \Lambda_q}
(k\Lambda + q\Lambda_q - p\Lambda_p) \, 
N^{\Lambda \Lambda_p \Lambda_q}_{-k \, p \, q} \, 
\langle 
{\cal Z}^{-s}_{\Lambda_p} {\cal Z}^s_{\Lambda_q} 
{\cal Z}^{s^{\prime}}_{\Lambda^{\prime}} 
\rangle
e^{-2 i s B_0 p_{\parallel} t} \, \delta_{pq,k} \, d{\bf p} \, d{\bf q} \, ,
$$
$$
+
$$
$$
\frac{\epsilon}{4}
\int \sum_{\Lambda_p, \Lambda_q}
(k^{\prime} \Lambda^{\prime} + q\Lambda_q - p\Lambda_p) \, 
N^{\Lambda^{\prime} \Lambda_p \Lambda_q}_{-k^{\prime} \, p \, q} \, 
\langle 
{\cal Z}^{-s^{\prime}}_{\Lambda_p} {\cal Z}^{s^{\prime}}_{\Lambda_q} 
{\cal Z}^s_{\Lambda}
\rangle
e^{-2 i s^{\prime} B_0 p_{\parallel} t} \, \delta_{pq,k^{\prime}} \, 
d{\bf p} \, d{\bf q} \, ,
$$
and 
\be
\partial_t \langle 
{\cal Z}^s_{\Lambda} {\cal Z}^{s^{\prime}}_{\Lambda^{\prime}} 
{\cal Z}^{s^{\prime\prime}}_{\Lambda^{\prime\prime}} 
\rangle = 
\label{thirdmhd}
\ee
$$
\frac{\epsilon}{4}
\int \sum_{\Lambda_p, \Lambda_q}
(k\Lambda + q\Lambda_q - p\Lambda_p) \, 
N^{\Lambda \Lambda_p \Lambda_q}_{-k \, p \, q} \, 
\langle 
{\cal Z}^{-s}_{\Lambda_p} {\cal Z}^s_{\Lambda_q} 
{\cal Z}^{s^{\prime}}_{\Lambda^{\prime}} 
{\cal Z}^{s^{\prime\prime}}_{\Lambda^{\prime\prime}} 
\rangle
e^{-2 i s B_0 p_{\parallel} t} \, \delta_{pq,k} \, d{\bf p} \, d{\bf q} \, ,
$$
$$
+
$$
$$
\frac{\epsilon}{4}
\int \sum_{\Lambda_p, \Lambda_q}
(k^{\prime} \Lambda^{\prime} + q\Lambda_q - p\Lambda_p) \, 
N^{\Lambda^{\prime} \Lambda_p \Lambda_q}_{-k^{\prime} \, p \, q} \, 
\langle 
{\cal Z}^{-s^{\prime}}_{\Lambda_p} {\cal Z}^{s^{\prime}}_{\Lambda_q} 
{\cal Z}^s_{\Lambda} {\cal Z}^{s^{\prime\prime}}_{\Lambda^{\prime\prime}} 
\rangle
e^{-2 i s^{\prime} B_0 p_{\parallel} t} \, \delta_{pq,k^{\prime}} \, 
d{\bf p} \, d{\bf q} \, ,
$$
$$
+
$$
$$
\frac{\epsilon}{4}
\int \sum_{\Lambda_p, \Lambda_q}
(k^{\prime\prime} \Lambda^{\prime\prime} + q\Lambda_q - p\Lambda_p) \, 
N^{\Lambda^{\prime\prime} \Lambda_p \Lambda_q}_{-k^{\prime\prime} \, p \, q} \, 
\langle 
{\cal Z}^{-s^{\prime\prime}}_{\Lambda_p} {\cal Z}^{s^{\prime\prime}}_{\Lambda_q} 
{\cal Z}^s_{\Lambda} {\cal Z}^{s^{\prime}}_{\Lambda^{\prime}} 
\rangle
e^{-2 i s^{\prime\prime} B_0 p_{\parallel} t} \, \delta_{pq,k^{\prime\prime}} \, 
d{\bf p} \, d{\bf q} \, . 
$$
We define the density tensor $q_{\Lambda\Lambda^{\prime}}^{s s^{\prime}}(\kk)$ 
for an homogeneous turbulence, 
\be
\langle 
{\cal Z}^{s}_{\Lambda} (\kk) \, {\cal Z}^{s^{\prime}}_{\Lambda^{\prime}} ({\kk}^{\prime})
\rangle \equiv q_{\Lambda\Lambda^{\prime}}^{s s^{\prime}}(\kk)
\, \delta (\kk + \kkpr) \, \delta_{s s^{\prime}} \, . 
\ee
The presence of the delta $\delta_{s s^{\prime}}$ means that correlations with opposite 
polarities have no long-time influence in the wave turbulence regime; the second delta 
distribution $\delta (\kk + \kkpr)$ is the consequence of the homogeneity assumption. 
We note that the kinematics does not impose any condition about the polarization 
$\Lambda$. The reason is that $\Lambda$ does not appear in the Alfv\'en wave frequency. 
This remark shows a fundamental difference with the Hall MHD case. As we will see it is 
the reason why principal value terms appear in MHD but not in Hall MHD. After the same 
kind of manipulations as in Appendix \ref{derivation}, the third-order moment 
equation (\ref{thirdmhd}) becomes:
\be
\langle 
{\cal Z}^s_{\Lambda} {\cal Z}^{s^{\prime}}_{\Lambda^{\prime}} 
{\cal Z}^{s^{\prime\prime}}_{\Lambda^{\prime\prime}} 
\rangle = 
\label{third21}
\ee
$$
\frac{\epsilon}{4} \, \delta_{k k^{\prime} k^{\prime \prime}}
\{ \, \sum_{\Lambda_p, \Lambda_q}
(k \Lambda + k^{\prime\prime} \Lambda_q - k^{\prime} \Lambda_p)
\left(N{\Lambda \Lambda_p \Lambda_q \atop k \, k^{\prime} \, k^{\prime \prime}}\right)^*
\Delta(2sB_0\kpa^{\prime}) \, 
q_{\Lambda_p \Lambda^{\prime}}^{-s-s} \, 
q_{\Lambda_q \Lambda^{\prime\prime}}^{ss}
$$
$$
+ 
$$
$$
\sum_{\Lambda_p, \Lambda_q}
(k \Lambda + k^{\prime} \Lambda_q - k^{\prime\prime} \Lambda_p)
\left(N{\Lambda \Lambda_p \Lambda_q \atop k \, k^{\prime\prime} \, k^{\prime}}\right)^*
\Delta(2sB_0\kpa^{\prime\prime}) \, 
q_{\Lambda_p \Lambda^{\prime\prime}}^{-s-s} \, 
q_{\Lambda_q \Lambda^{\prime}}^{ss}
$$
$$
+
$$
$$
\sum_{\Lambda_p, \Lambda_q}
(k^{\prime} \Lambda^{\prime} + k^{\prime\prime} \Lambda_q - k \Lambda_p)
\left(N{\Lambda^{\prime} \Lambda_p \Lambda_q \atop k^{\prime} \, k \, k^{\prime \prime}}
\right)^*
\Delta(2s^{\prime}B_0\kpa) \, 
q_{\Lambda_p \Lambda}^{-s^{\prime}-s^{\prime}} \, 
q_{\Lambda_q \Lambda^{\prime\prime}}^{s^{\prime}s^{\prime}}
$$
$$
+ 
$$
$$
\sum_{\Lambda_p, \Lambda_q}
(k^{\prime} \Lambda^{\prime} + k \Lambda_q - k^{\prime\prime} \Lambda_p)
\left(N{\Lambda^{\prime} \Lambda_p \Lambda_q \atop k^{\prime} \, k^{\prime \prime} \, k}
\right)^*
\Delta(2s^{\prime}B_0\kpa^{\prime\prime}) \, 
q_{\Lambda_p \Lambda^{\prime\prime}}^{-s^{\prime}-s^{\prime}} \, 
q_{\Lambda_q \Lambda}^{s^{\prime}s^{\prime}}
$$
$$
+
$$
$$
\sum_{\Lambda_p, \Lambda_q}
(k^{\prime\prime} \Lambda^{\prime\prime} + k^{\prime} \Lambda_q - k \Lambda_p)
\left(N{\Lambda^{\prime\prime} \Lambda_p \Lambda_q \atop k^{\prime\prime} \, k \, 
k^{\prime}}\right)^*
\Delta(2s^{\prime\prime}B_0\kpa) \, 
q_{\Lambda_p \Lambda}^{-s^{\prime\prime}-s^{\prime\prime}} \, 
q_{\Lambda_q \Lambda^{\prime}}^{s^{\prime\prime}s^{\prime\prime}}
$$
$$
+
$$
$$
\sum_{\Lambda_p, \Lambda_q}
(k^{\prime\prime} \Lambda^{\prime\prime} + k \Lambda_q - k^{\prime} \Lambda_p)
\left(N{\Lambda^{\prime\prime} \Lambda_p \Lambda_q \atop k^{\prime\prime} \, k^{\prime} 
\, k}\right)^*
\Delta(2s^{\prime\prime}B_0\kpa^{\prime} ) \, 
q_{\Lambda_p \Lambda^{\prime}}^{-s^{\prime\prime}-s^{\prime\prime}} \, 
q_{\Lambda_q \Lambda}^{s^{\prime\prime}s^{\prime\prime}}
\} \, . 
$$
We insert expression (\ref{third21}) into (\ref{secondmhd}). We note that only the third 
and the fifth terms of expression (\ref{third21}) will contribute. One obtains
\be
\partial_t 
q_{\Lambda\Lambda^{\prime}}^{s s}(\kk) = 
\label{secondmhdbis}
\ee
$$
\frac{\epsilon}{16} \{ \, 
\int \sum_{\Lambda_p, \Lambda_q \atop {\bar \Lambda_p}, {\bar \Lambda_q}}
(k\Lambda + q\Lambda_q - p\Lambda_p) \, (q\Lambda_q + k {\bar \Lambda_q}
- p {\bar \Lambda_p}) \, 
N{\Lambda \Lambda_p \Lambda_q \atop -k \, p \, q}
\left(-N{{\bar \Lambda_q} {\bar \Lambda_p} \Lambda_q \atop -k \, p \, q}\right)^*
$$
$$
\Delta(-2sB_0p_{\parallel}) \, q_{{\bar \Lambda_p} \Lambda_p}^{-s -s} \, 
q_{{\bar \Lambda_q} \Lambda^{\prime}}^{ss} \, 
\delta_{pq,k} \, d{\bf p} \, d{\bf q}
$$
$$
+
$$
$$
\int \sum_{\Lambda_p, \Lambda_q \atop {\bar \Lambda_p}, {\bar \Lambda_q}}
(k\Lambda + q\Lambda_q - p\Lambda_p) \, (k \Lambda^{\prime} +
q {\bar \Lambda_q} - p {\bar \Lambda_p}) \, 
N{\Lambda \Lambda_p \Lambda_q \atop -k \, p \, q}
\left(N{\Lambda {\bar \Lambda_p} {\bar \Lambda_q} \atop -k \, p \, q}\right)^*
$$
$$
\Delta(-2sB_0p_{\parallel}) \, q_{{\bar \Lambda_p} \Lambda_p}^{-s -s} \, 
q_{{\bar \Lambda_q} \Lambda_q}^{ss} \, 
\delta_{pq,k} \, d{\bf p} \, d{\bf q}
$$
$$
+
$$
$$
\int \sum_{\Lambda_p, \Lambda_q \atop {\bar \Lambda_p}, {\bar \Lambda_q}}
(k \Lambda^{\prime} + q\Lambda_q - p\Lambda_p) \, 
(q\Lambda_q + k {\bar \Lambda_q} - p {\bar \Lambda_p}) \, 
N{\Lambda^{\prime} \Lambda_p \Lambda_q \atop k \, p \, q}
\left(-N{{\bar \Lambda_q} {\bar \Lambda_p} \Lambda_q \atop k \, p \, q}\right)^*
$$
$$
\Delta(-2sB_0p_{\parallel}) \, q_{{\bar \Lambda_p} \Lambda_p}^{-s -s} \, 
q_{{\bar \Lambda_q} \Lambda}^{ss} \, 
\delta_{kpq} \, d{\bf p} \, d{\bf q}
$$
$$
+
$$
$$
\int \sum_{\Lambda_p, \Lambda_q \atop {\bar \Lambda_p}, {\bar \Lambda_q}}
(k \Lambda^{\prime} + q\Lambda_q - p\Lambda_p) \, 
(k \Lambda + q{\bar \Lambda_q} - p {\bar \Lambda_p}) \, 
N{\Lambda^{\prime} \Lambda_p \Lambda_q \atop k \, p \, q}
\left(N{\Lambda {\bar \Lambda_p} {\bar \Lambda_q} \atop k \, p \, q}\right)^*
$$
$$
\Delta(-2sB_0p_{\parallel}) \, q_{{\bar \Lambda_p} \Lambda_p}^{-s -s} \, 
q_{{\bar \Lambda_q} \Lambda_q}^{ss} \, 
\delta_{kpq} \, d{\bf p} \, d{\bf q} \, \} \, .
$$
We change signs for wavevectors $\pp$ and $\qq$ in the last two terms and 
obtain:
\be
\partial_t 
q_{\Lambda\Lambda^{\prime}}^{s s}(\kk) = \frac{\epsilon}{16} 
\int \sum_{\Lambda_p, \Lambda_q \atop {\bar \Lambda_p}, {\bar \Lambda_q}} 
\label{secondmhdter}
\ee
$$
\{ \, (k\Lambda + q\Lambda_q - p\Lambda_p) \, (q\Lambda_q + k {\bar \Lambda_q}
- p {\bar \Lambda_p}) \, 
N{\Lambda \Lambda_p \Lambda_q \atop -k \, p \, q}
\left(-N{{\bar \Lambda_q} {\bar \Lambda_p} \Lambda_q \atop -k \, p \, q}\right)^*
\Delta(-2sB_0p_{\parallel}) \, q_{{\bar \Lambda_p} \Lambda_p}^{-s -s} \, 
q_{{\bar \Lambda_q} \Lambda^{\prime}}^{ss} \, 
$$
$$
+
$$
$$
(k\Lambda + q\Lambda_q - p\Lambda_p) \, (k \Lambda^{\prime} +
q {\bar \Lambda_q} - p {\bar \Lambda_p}) \, 
N{\Lambda \Lambda_p \Lambda_q \atop -k \, p \, q}
\left(N{\Lambda {\bar \Lambda_p} {\bar \Lambda_q} \atop -k \, p \, q}\right)^*
\Delta(-2sB_0p_{\parallel}) \, q_{{\bar \Lambda_p} \Lambda_p}^{-s -s} \, 
q_{{\bar \Lambda_q} \Lambda_q}^{ss} \, 
$$
$$
+
$$
$$
(k \Lambda^{\prime} + q\Lambda_q - p\Lambda_p) \, 
(q\Lambda_q + k {\bar \Lambda_q} - p {\bar \Lambda_p}) \, 
N{{\bar \Lambda_q} {\bar \Lambda_p} \Lambda_q \atop -k \, p \, q}
\left(-N{\Lambda^{\prime} \Lambda_p \Lambda_q \atop -k \, p \, q}\right)^*
\Delta(2sB_0p_{\parallel}) \, q_{{\bar \Lambda_p} \Lambda_p}^{-s -s} \, 
q_{{\bar \Lambda_q} \Lambda}^{ss} \, 
$$
$$
+
$$
$$
(k \Lambda^{\prime} + q\Lambda_q - p\Lambda_p) \, 
(k \Lambda + q{\bar \Lambda_q} - p {\bar \Lambda_p}) \, 
N{\Lambda {\bar \Lambda_p} {\bar \Lambda_q} \atop -k \, p \, q}
\left(N{\Lambda^{\prime} \Lambda_p \Lambda_q \atop -k \, p \, q}\right)^*
\Delta(2sB_0p_{\parallel}) \, q_{{\bar \Lambda_p} \Lambda_p}^{-s -s} \, 
q_{{\bar \Lambda_q} \Lambda_q}^{ss} \, \}
$$
$$
\delta_{pq,k} \, d{\bf p} \, d{\bf q} \, .
$$
The long-time behavior is then given by the Riemman-Lebesgue Lemma:
\be
\Delta(\pm 2 sB_0p_{\parallel} ) \to 
\pi \delta(2 sB_0p_{\parallel}) \mp i {\cal P}(1/2 sB_0p_{\parallel}) \, .
\ee
We see that principal value terms will appear in the long-time limit. The reason 
is that the polarization $\Lambda$ and $\Lambda^{\prime}$ in the density tensor 
$q_{\Lambda\Lambda^{\prime}}^{s s}(\kk)$ are not the same in general. Therefore
we lose the symmetry between terms that we had in the Hall MHD case (see Appendix
\ref{derivation}).

\section{Simplified MHD wave kinetic equations}
\label{AppenE}

In in section, we continue the analysis made in Appendix \ref{AppenD} when only 
terms symmetric in $\Lambda$ are retained, \ie terms like $q_{\Lambda \Lambda}^{ss}$. 
Then expression (\ref{secondmhdter}) simplifies to
\be
\partial_t 
q_{\Lambda\Lambda}^{s s}(\kk) = \frac{\epsilon}{16} 
\int \sum_{\Lambda_p, \Lambda_q} 
\label{E1}
\ee
$$
\{ \, (k\Lambda + q\Lambda_q - p\Lambda_p)^2 \, 
N{\Lambda \Lambda_p \Lambda_q \atop -k \, p \, q}
\left(-N{\Lambda \Lambda_p \Lambda_q \atop -k \, p \, q}\right)^*
\Delta(-2sB_0p_{\parallel}) \, q_{\Lambda_p \Lambda_p}^{-s -s} \, 
q_{\Lambda \Lambda}^{ss} \, 
$$
$$
+
$$
$$
(k\Lambda + q\Lambda_q - p\Lambda_p)^2 \, 
\, N{\Lambda \Lambda_p \Lambda_q \atop -k \, p \, q}
\left(N{\Lambda \Lambda_p \Lambda_q \atop -k \, p \, q}\right)^*
\Delta(-2sB_0p_{\parallel}) \, q_{\Lambda_p \Lambda_p}^{-s -s} \, 
q_{\Lambda_q \Lambda_q}^{ss} \, 
$$
$$
+
$$
$$
(k \Lambda + q\Lambda_q - p\Lambda_p)^2 \, 
N{\Lambda \Lambda_p \Lambda_q \atop -k \, p \, q}
\left(-N{\Lambda \Lambda_p \Lambda_q \atop -k \, p \, q}\right)^*
\Delta(2sB_0p_{\parallel}) \, q_{\Lambda_p \Lambda_p}^{-s -s} \, 
q_{\Lambda \Lambda}^{ss} \, 
$$
$$
+
$$
$$
(k \Lambda + q\Lambda_q - p\Lambda_p)^2 \, 
N{\Lambda \Lambda_p \Lambda_q \atop -k \, p \, q}
\left(N{\Lambda \Lambda_p \Lambda_q \atop -k \, p \, q}\right)^*
\Delta(2sB_0p_{\parallel}) \, q_{\Lambda_p \Lambda_p}^{-s -s} \, 
q_{\Lambda_q \Lambda_q}^{ss} \}
$$
$$
\delta_{pq,k} \, d{\bf p} \, d{\bf q} \, .
$$
The symmetry between terms is recovered; further simplifications lead to:
\be
\partial_t 
q_{\Lambda\Lambda}^{s s}(\kk) = \frac{\epsilon}{16} 
\int \sum_{\Lambda_p, \Lambda_q} 
\left|N{\Lambda \Lambda_p \Lambda_q \atop -k \, p \, q}\right|^2
(k\Lambda + q\Lambda_q - p\Lambda_p)^2 \, 
\label{E2}
\ee
$$
( \Delta(-2sB_0p_{\parallel}) + \Delta(2sB_0p_{\parallel}) )
\, q_{\Lambda_p \Lambda_p}^{-s -s} \, 
( q_{\Lambda_q \Lambda_q}^{ss} - q_{\Lambda \Lambda}^{ss} ) \, 
\delta_{pq,k} \, d{\bf p} \, d{\bf q} \, .
$$
The long-time behavior is given by the Riemman-Lebesgue Lemma; one finds:
\be
\partial_t 
q_{\Lambda\Lambda}^{s s}(\kk) = 
\label{E3}
\ee
$$
\frac{\pi \epsilon}{16 B_0} 
\int \sum_{\Lambda_p, \Lambda_q} 
\left( \frac{\sin \psi_k}{k} \right)^2 
(k\Lambda + p\Lambda_p + q\Lambda_q)^2 \, (k\Lambda + q\Lambda_q - p\Lambda_p)^2 
$$
$$
q_{\Lambda_p \Lambda_p}^{-s -s} \, 
( q_{\Lambda_q \Lambda_q}^{ss} - q_{\Lambda \Lambda}^{ss} ) \, 
\delta(p_{\parallel}) \, \delta_{k,pq} \, d{\bf p} \, d{\bf q} \, .
$$
Equations (\ref{E3}) are the wave kinetic equations for incompressible MHD
turbulence at the level of three-wave interactions when helicities are absent 
and when equality between shear- and pseudo-Alfv\'en wave energies is assumed
(see Section \ref{45}). 
We find the same equations as (\ref{ke_alfven2}) since we have the relation
$q_{\Lambda\Lambda}^{s s}(\kk) = 4 \, \qls (\kk)$.

\section{Kinetic equations for the energies $E$ and $E_d$}
\label{AppenC}

We introduce the expression (\ref{qexpand}) into (\ref{ke1}) and we consider a state 
of zero helicities ($H_m(\kk)=0$ and $H_G(\kk)=0$). The kinetic equations for the 
energies $E$ and $E_d$ are 
\be
\partial_t {E (\kk) \brace E_d(\kk)} = 
\label{masters}
\ee
$$
\frac{\pi \, \epsilon^2}{32 \, d_i^2 B_0^2} \int 
\sum_{\Lambda, \Lambda_p, \Lambda_q \atop s, s_p, s_q} 
\left( \frac{\sin \psi_k}{k} \right)^2 \, 
\frac{(\Lambda k + \Lambda_p p + \Lambda_q q)^2 
\, \left( 1 - {\alfm}^2 {\alfmp}^2 {\alfmq}^2 \right)^2}
{(1 + {\alfm}^2)(1 + {\alfmp}^2)(1 + {\alfmq}^2)}
$$
$$
\left( \frac{\alfq - \alfp}{\kpa} \right)^2 \, 
{\alfm+1 \brace \alfm-1} \, \frac{\ols \, \olsp}{{\alfm}^2+1}
\left[
E(\qq) + \left(\frac{{\alfmq}^2+1}{{\alfmq}^2-1}\right) E_d(\qq) 
\right]
$$
$$
\left[ 
E(\pp) - E(\kk) + \left(\frac{{\alfmp}^2+1}{{\alfmp}^2-1}\right) E_d(\pp)
- \left(\frac{{\alfm}^2+1}{{\alfm}^2-1}\right) E_d(\kk) 
\right] \delta(\Omega_{k,pq})\, \delta_{k,pq} \, d{\bf p} \, d{\bf q} \, . 
$$

\section{Compatibility with Galtier et al. (2000)}
\label{AppenF}

We start from equation (26) of Galtier et al. (2000). We assume that helicities
are absent ($I^s(\kk)=0$) and that shear- and pseudo-Alfv\'en wave energies are 
identical. We introduce the following notations:
\be
\kpn^2 \Psi^s(\kk) = e^s(\kk) / 2 \, ,
\ee
and
\be
\kpn^2 k^2 \Phi^s(\kk) = e^s(\kk) / 2 \, .
\ee
Then equation (26) writes:
\be
\partial_t e^s(\kk) =
\label{F1}
\ee
$$
\frac{\pi \, \epsilon^2}{4 B_0} \int \{ 
(1 - \frac{(\kk \times \kappa)^2}{L_{\perp}^2 k^2} + 
\frac{\kpa^2 \kappa^2_{\perp}}{L_{\perp}^2k^2}) \, e^s({\bf L})
- (1 - \frac{(\kk \times \kappa)^2}{\kpn^2 L^2} + 
\frac{\kpa^2 \kappa^2_{\perp}}{\kpn^2 L^2}) \, e^s(\kk)
$$
$$
+
$$
$$
(1 - \frac{\kpa^2 \kappa^2 \cos^2 \psi_k}{L_{\perp}^2 k^2}) \, e^s({\bf L}) 
- (1 - \frac{\kpa^2 \kappa^2 \cos^2 \psi_L}{\kpn^2 L^2}) \, e^s(\kk) \, \}
$$
$$
\{ \, (\kk \times \kappa)^2 - \kpa^2 \kappa_{\perp}^2) 
\frac{e^{-s}(\kappa)}{\kappa_{\perp}^2} + \kpa^2 \, e^{-s}(\kappa) \} \, 
\delta(\kappa_{\parallel}) \, \delta_{k,\kappa L} d\kappa d{\bf L} \, .
$$
We remind that the angle $\psi_k$ refers to the angle opposite to the 3D vector 
$\kk$ in the triangle defined by $\kk={\bf L}+\kappa$. (In Galtier et al. 
(2000), angles are introduced in reference to 2D wavevectors.) 
Some manipulations lead to:
\be
\partial_t e^s(\kk) =
\label{F2}
\ee
$$
\frac{\pi \, \epsilon^2}{4 B_0} \int \{ 
(2 - \frac{\kappa_{\perp}^2 \sin^2 \psi_L}{L_{\perp}^2} + 
\frac{\kpa^2 \kappa^2_{\perp}}{L_{\perp}^2 k^2} - 
\frac{\kpa^2 \kappa_{\perp}^2 \cos^2 \psi_k}{L_{\perp}^2 k^2}) \, e^s({\bf L})
$$
$$
-
$$
$$
(2 - \frac{(k^2 \kappa_{\perp}^2 \sin^2 \psi_L}{\kpn^2 L^2} + 
\frac{\kpa^2 \kappa^2_{\perp}}{\kpn^2 L^2} - 
\frac{\kpa^2 \kappa_{\perp}^2 \cos^2 \psi_L}{\kpn^2 L^2} ) \, e^s(\kk) \}
$$
$$
k^2 \sin^2 \psi_L \, e^{-s}(\kappa) \, 
\delta(\kappa_{\parallel}) \, \delta_{k,\kappa L} d\kappa d{\bf L} \, ,
$$
that can be written as:
\be
\partial_t e^s(\kk) =
\label{F3}
\ee
$$
\frac{\pi \, \epsilon^2}{4 B_0} \int \{ 
(1 + \cos^2 \psi_{\kappa}) \, L^2 \, \sin^2 \psi_k \, e^{-s}(\kappa) \, 
(e^s({\bf L}) - e^s(\kk)) \, \delta(\kappa_{\parallel}) \, \delta_{k,\kappa L} 
d\kappa d{\bf L} \, .
$$
Equations (\ref{F3}) are the wave kinetic equations for incompressible MHD
turbulence when helicities are absent and when equality between shear- and 
pseudo-Alfv\'en wave energies is assumed. The addition over the index $s$ will 
give the wave kinetic equations for the total energy.

\section{Nonlocal WCC interactions}
\label{AppenG}

We shall consider the strongly nonlocal interactions of two ion cyclotron waves at small 
scales on one whistler wave at large scales. In other words, it means that the whistler 
wave will be supported by the wavevector $\kk$, with $k \ll p, q$. This type of interaction 
is the most interesting since smaller scales may be reached more easily by ion cyclotron 
waves. For such a limit, the master equations (\ref{mastersHall}) reduce to
\be
\partial_t {E^V (\kk) \brace E^B(\kk)} = 
\label{WWA2}
\ee
$$
\frac{\pi \, \epsilon^2}{8 \, d_i^4} \int \sum_{s, s_p, s_q} 
\left( \frac{\sin \psi_k}{k} \right)^2 \, 
\left( \frac{s_q /q - s_p /p}{\kpa} \right)^2 \, 
(s_p p + s_q q)^2 \, \frac{p q^2 \kpa p_{\parallel}}{k^3} \, {\frac{1}{d_i^2 k^2} \brace 1}
$$
$$
E^V(\qq) \left[ E^V(\pp) - E^B(\kk) \right] 
\delta(\Omega_{k,pq})\, \delta_{k,pq} \, d{\bf p} \, d{\bf q} \, . 
$$
As expected, at leading order the whistler wave is described by the magnetic energy. 
Contrary to the other cases studied before, a non trivial dynamics happens as long as 
a discrepancy exists between the (kinetic) energy of the ion cyclotron waves and the 
(magnetic) energy of the whistler wave.

\end{document}